\documentclass{aastex631}

\usepackage{mhchem}
\usepackage{rotating}
\usepackage{booktabs}

\shorttitle{HNCO for outflows}
\shortauthors{Xie et al.}
\graphicspath{{./}{figures/}}

\begin{document}

\title{Imaging Molecular Outflow in Massive Star-forming Regions with HNCO Lines}

\correspondingauthor{Juan Li, Junzhi Wang, Jinjin Xie}
\email{lijuan@shao.ac.cn, junzhiwang@gxu.edu.cn, jinjinxie@shao.ac.cn}

\author[0000-0002-2738-146X]{Jinjin Xie}
\affiliation{Shanghai Astronomical Observatory, 80 Nandan Road, Shanghai 200030, China}

\author[0000-0003-3520-6191]{Juan Li}
\affiliation{Shanghai Astronomical Observatory, 80 Nandan Road, Shanghai 200030, China}


\author[0000-0001-6106-1171]{Junzhi Wang}
\affiliation{Guangxi Key Laboratory for Relativistic Astrophysics, Department of Physics, Guangxi University, Nanning 530004, PR China}
\author[0000-0001-6016-5550]{Shu Liu}
\affiliation{National Astronomical Observatories, Chinese Academy of Sciences, A20 Datun Road, Chaoyang District, Beijing 100101, China}

\author[0000-0002-9839-185X]{Kai Yang}
\affiliation{School of Astronomy and Space Science, Nanjing University, 163 Xianlin Avenue, Nanjing 210023, China}

\author[0000-0003-4811-2581]{Donghui Quan}
\affiliation{Research Center for Intelligent Computing Platforms, Zhejiang Laboratory, Hangzhou 311100, China} 

\author[0000-0001-9047-846X]{Siqi Zheng}
\affiliation{Shanghai Astronomical Observatory, 80 Nandan Road, Shanghai 200030, China}
\affiliation{University of Chinese Academy of Sciences, Beijing 100049, China}

\author{Yuqiang Li}
\affiliation{Shanghai Astronomical Observatory, 80 Nandan Road, Shanghai 200030, China}
\affiliation{University of Chinese Academy of Sciences, Beijing 100049, China}

\author{Jingwen Wu}
\affiliation{University of Chinese Academy of Sciences, Beijing 100049, China}

\author[0000-0003-3758-7426]{Yan Duan}
\affiliation{National Astronomical Observatories, Chinese Academy of Sciences, A20 Datun Road, Chaoyang District, Beijing 100101, China}
\affiliation{University of Chinese Academy of Sciences, Beijing 100049, China}

\author[0000-0003-3010-7661]{Di Li}
\affiliation{National Astronomical Observatories, Chinese Academy of Sciences, A20 Datun Road, Chaoyang District, Beijing 100101, China}
\affiliation{Research Center for Intelligent Computing Platforms, Zhejiang Laboratory, Hangzhou 311100, China}
\affiliation{NAOC-UKZN Computational Astrophysics Centre, University of KwaZulu-Natal, Durban 4000, South Africa}





\begin{abstract}

Protostellar outflows are considered a signpost of star formation. These outflows can cause shocks in the molecular gas and are typically traced by the line wings of certain molecules. HNCO (4--3) has been regarded as a shock tracer because of the high abundance in shocked regions. Here we present the first imaging results of HNCO (4--3) line wings toward nine sources in a sample of twenty three massive star-forming regions using the IRAM 30\,m telescope. We adopt the velocity range of the full width of HC$_{3}$N (10--9) and H$^{13}$CO$^+$ (1--0) emissions as the central emission values, beyond which the emission from HNCO (4--3) is considered to be from line wings. The spatial distributions of the red- and/or blue-lobes of HNCO (4--3) emission nicely associate with those lobes of HCO$^{+}$ (1--0) in most of the sources. High intensity ratios of HNCO (4--3) to HCO$^+$ (1--0) are obtained in the line wings. The derived column density ratios of HNCO to HCO$^+$ are consistent with those previously observed towards massive star-forming regions. These results provide direct evidence that HNCO could trace outflow in massive star-forming regions. This work also implies that the formation of some HNCO molecules is related to shock, either on the grain surface or within the shocked gas.


\end{abstract}



\section{Introduction} \label{sec:intro}

Massive star-forming regions are considered to be exposed to shocks, which are powered by cloud collisions, infall motions, OB stellar winds, and ionization fronts \citep{2018motte}. Shocks including outflows are supposed to be more vigorous in the later stages of massive star formation. Outflows usually manifest as broad line wings in spectra and spatially separated red- and/or blue-lobes centered on young stellar objects, and often show increased abundances in shock tracers such as SiO, SO, CH$_{3}$OH, NH$_{3}$, and Isocyanic acid (HNCO) \citep[e.g.][]{1993bachiller,1994chernin,2007kalenskii,2010rodriguezfernandez,2019lishanghuo}.

Among these shock tracers, HNCO is peculiar both chemically and astrophysically. Chemically, HNCO is a nearly prolate asymmetric top molecule and the simplest complex organic species containing C, H, O, and N elements \citep{1950jones}. Resembling a peptide bond, HNCO has been considered among the precursors of prebiotic molecules \citep{2007bisschop,2018ferus,2018quenard}. Since its first detection in Sagittarius B2 \citep{1971snyder}, HNCO has been ubiquitously detected in various interstellar circumstances, from comets to external galaxies \citep[e.g.][]{1981brown,1991nguyenqrieu,1997lis,2015velillaprieto}. Observations of HNCO towards dense hot clumps associated with protostars and H\uppercase\expandafter{\romannumeral2}  regions \citep[e.g.][]{2013lijuan,2013jackson} indicated that HNCO might be related with star formation. The role of tracing shocks has been suggested for HNCO from its associations with other shock tracers, e.g., seen in the integrated intensities maps \citep{2000zinchenko,2010rodriguezfernandez,2012sanhueza,2014miettinen}. HNCO spectra with broad line wings have been noticed towards low mass \citep{2010rodriguezfernandez} and massive star-forming regions \citep{2000zinchenko,2013lijuan,2021canelo}. However, spatially resolved analyses on the line wings of HNCO, which may be from outflows, are still lacking.

The high abundance of HNCO in the shocked regions has been considered to support that HNCO is a shock tracer. The abundances of HNCO are typically within the range of 10$^{-10}$ to 10$^{-8}$ \citep[e.g.][]{2000zinchenko,2007bisschop,2010quan}. The highest value 1$\times$10$^{-7}$ of HNCO abundance was detected towards the shocked region of L1157 \citep{2010rodriguezfernandez}, which is a well-studied low-mass star-forming region \citep[e.g.][]{1997bachiller,2020james}. From the astrochemical prospects, the abundance of HNCO could increase in warm dense gas \citep{2017kelly}. Most recent chemical modelings introduced in shocks indicated that the abundance of HNCO can be increased by one order of magnitude even in weakly shocked regions \citep{2020zhangxia}. Thus, substantial HNCO intensities should be observed towards shocked regions in massive star formation.

In this paper, we present the observational data from IRAM 30\,m telescope towards a sample of 9 hot cores at the later stages of massive star formation. We provide the results of the line wings, the outflow lobes, and the {high intensity ratios of HNCO to HCO$^+$, further supporting HNCO as a shock tracer. Sect. \ref{sec:observation} explains the observational parameters. Sect. \ref{sec:results} presents the mapping results of the HNCO line wing and the derived intensity ratios of HNCO (4--3) to HCO$^+$ (1--0) in the line wings. Comments on individual sources are given in Sect. \ref{subsec:disc2}. The results are discussed in Sect. \ref{sec:discussion}. We summarize the results in Sect. \ref{ref:summary}.

\section{Observation and Data Reduction} \label{sec:observation}

The sources are selected from a sample of 23 massive star-forming regions (Liu et al. in prep.), which have been measured trigonometric parallaxes with class II methanol masers or water masers \citep{2014reid} and show spatial distributions in H$^{13}$CN (3--2) (Liu et al. in prep.). The observed sources are listed in Table ~\ref{tab:parameters}. The observations were done with the Instituto de Radioastronom\'ia Milim\'etrica (IRAM) 30-m telescope at Pico Veleta, Spain in July, October, and November 2019, December 2020, and January 2021. The data were taken with the 3 mm (E0) band of the Eight Mixer Receiver (EMIR) and the Fourier Transform Spectrometers (FTS) backend to cover 8 GHz bandwidth with a 195 kHz spectral resolution for each band in dual polarization. 
The beam size of IRAM 30\,m is $\sim$ 27$^{''}$ at 90\,GHz. The typical system temperatures were around 150 K in the 3\,mm band. Pointing was checked every 2 hours with nearby strong quasi-stellar objects. Focus was checked and corrected at the beginning of each run and during sunsets/sunrises. The antenna temperature ($T_{\rm A}^{\ast}$) was converted to the main beam brightness temperature ($T_{\rm mb}$) using $T_{\rm mb}$=$T_{\rm A}^{\ast}\cdot F_{\rm eff}/B_{\rm eff}$, where the forward efficiency $F_{\rm eff}$ is 0.95 and beam efficiency $B_{\rm eff}$ is 0.81 for 3 mm band. Mapping observations were carried out using the On-The-Fly (OTF) mode. The off-positions were -600$''$ to the peak of each mapping region. Each scan consists of 2 minutes with an on-source integration of 1 minute. We used the {\sc CLASS} package, which is a part of the {\sc GILDAS}\footnote{http://www.iram.fr/IRAMFR/GILDAS} software, to reduce the data. The spectra were smoothed to 1.33\,km s$^{-1}$ to improve the signal-to-noise ratio of the measured line emission. The pixel size was resampled to 9$''$ for each source.

We observed HNCO ($J_{Ka,Kb}$=4$_{0,4}$-3$_{0,3}$) transition (hereafter HNCO (4--3)) at 87925.238\,MHz. HCO$^{+}$ J=1--0 (hereafter HCO$^+$(1--0)) at 89188.526\,MHz and its optically thin isotope H$^{13}$CO$^{+}$ J=1--0 (hereafter H$^{13}$CO$^+$(1--0)) at 86754.288\,MHz are used to distinguish outflow. The observation also covered HC$_{3}$N J=10--9 (hereafter HC$_{3}$N(10--9)) at 90978.989\,MHz for tracing dense gas. The H\uppercase\expandafter{\romannumeral2}  region is represented with radio recombination line H42$\alpha$ at 85688.39\,MHz, which traces ionized gas. The physical parameters for the observed lines are listed in Table \ref{tab:calculation}.

\section{Results}\label{sec:results}

\subsection{Line wings of HNCO (4--3) emission}\label{subsec:linewing}

Fig. \ref{fig:spec} and Fig. \ref{fig:spec_hcop} show the spectra of the observed lines averaged from the HNCO (4--3) contoured red- and/or blue-shifted emission areas, which are labeled in Fig. \ref{fig:outflow} and Fig. \ref{fig:bluered}. The 3$\sigma$ noise level is $\sim$0.2\,K at 1.33 km s$^{-1}$ velocity resolution. HNCO(4--3) line wings in most sources could be distinguished from Fig. \ref{fig:spec} and Fig. \ref{fig:spec_hcop}. The velocity range of the line wing is determined from the whole emission above zero intensity excluding the line center emission, which is identified by visually inspecting the spectra of the optically thin lines H$^{13}$CO$^+$( 1--0) and HC$_{3}$N (10--9). The red and/or blue line wings are labeled with slashed windows in corresponding colors, as shown in Fig. \ref{fig:spec}. Fig. \ref{fig:spec} also includes the spectra of the two optically thin lines. In four sources, G015.03-00.67, G075.76+00.33, G109.87+02.11, and G111.54+00.77, only blue line wings are prominent. In the northern region of source G034.39+00.22, only red line wings are prominent. The rest of the sources show both red and blue line wings. For sources that lack either red- or blue-line wing, the velocity range of the line core is labeled with the black window. 

HCO$^+$ (1--0), a commonly used outflow/infall tracer, is also used to help distinguish outflows. The spectra of HCO$^+$ (1--0) together with the two optically thin lines are included in Fig. \ref{fig:spec_hcop}. HCO$^+$ (1--0) shows clear line wings in all sources. In the majority of the sources, the preferences for red- and/or blue-shifted line wings in HCO$^+$ (1--0) are consistent with those in HNCO (4--3), except for G109.87+02.11 and G111.54+00.77.  The velocity ranges of HNCO (4--3) and HCO$^+$ (1--0) line wings are listed in Table~\ref{tab:range}. In most of the sources, the velocity range of the line wing of HNCO (4--3) is equivalent to or slightly less than that of HCO$^+$ (1--0). Thus, the line wings of HNCO (4--3) are attributed to outflows.

\subsection{Integrated intensities of HNCO line wings }\label{subsec:lobe}

The intensities of HNCO (4--3) line wings are integrated to depict the outflow lobes. The contours for the integrated intensities of the molecular lines in all sources shown in Fig. \ref{fig:outflow} and Fig. \ref{fig:bluered} are drawn from 3$\sigma$ level with the increment of 2$\sigma$ if not specified in the captions of the figures. The integrated intensities of HNCO (4--3) towards the line core and line wings, the integrated intensities of H$^{13}$CO$^+$ (1--0) towards the line core, and the integrated intensities of HCO$^+$ (1--0) towards the line wings are listed in Table~\ref{tab:range}. The distributions of HNCO (4--3) red- and/or blue-lobes are similar to those of HCO$^+$ (1--0) in most of the sources, except G111.54+00.77. Fig. \ref{fig:outflow} presents the three sources showing both red- and blue-shifted emissions of HNCO (4--3). Fig. \ref{fig:bluered} includes all other sources. The centers of the bipolar outflows traced by HNCO (4--3) and HCO$^+$ (1--0) are not always at the dense cores traced by HC$_{3}$N (10--9). In G075.76+00.33 and G192.60-00.04, the red- or blue-shifted emissions of HNCO (4--3) integrated intensities are distributed adjacent to H\uppercase\expandafter{\romannumeral2} regions.

  \subsection{Intensity ratios and column density ratios of HNCO to HCO$^+$}\label{subsec:abundance}

With the assumptions that the excitation conditions are similar in the line center and in the outflowing gas, the intensity ratios of HNCO (4--3) to HCO$^+$ (1--0) ($I[HNCO]/I[HCO^+]$, [HNCO]/[HCO$^+$] hereafter) in the line center and the outflowing gas can be compared. The velocity ranges of HNCO(4--3) and HCO$^+$(1--0) line center and line wings are listed in Table~\ref{tab:range}, which also includes the corresponding integrated intensities of HNCO (4--3) and HCO$^+$ (1--0). From Table~\ref{tab:range}, the intensity ratios of HNCO (4--3) to HCO$^+$ (1--0) in the line center range from (0.28$\pm$0.01)\% to (1.55$\pm$0.06)\%, with the median of (0.53$\pm$0.05)\%. The ranges of the intensity ratios of HNCO (4--3) to HCO$^+$ (1--0) are from 9.37$\pm$1.95)\% to (23.34$\pm$2.33)\%, with the median of (11.29$\pm$1.58)\% in the red lobes, and are from (2.97$\pm$0.66)\% to (36.49$\pm$7.82)\%, with the median of (11.07$\pm$3.17)\% in the blue lobes.

The column density ratios of HNCO to HCO$^+$ could be converted from the intensity ratios of HNCO (4--3) to HCO$^+$ (1--0) assuming the molecular lines are optically thin. Under the optically thin and LTE assumptions, the column density is calculated from
  \begin{eqnarray}\label{eq:ndensity}
N_{tot}&=&\frac{8\pi}{\lambda^3 A}\frac{g_l}{g_i}\frac{1}{J_{\nu}\left(T_{ex}\right)-J_{\nu}\left(T_{bg}\right)}\frac{1}{1-exp\left(-h\nu/kT_{tex}\right)}\times\frac{Q_{rot}}{g_{l}exp\left(-E_{l}/kT_{ex}\right)} \int T_{mb}dv,
\end{eqnarray}
\citep{2011vasyunina}. $\lambda$ is the rest wavelength of the transition, $\lambda \equiv c/\nu$. $A$ is the Einstein spontaneous coefficient. $g_{u}$, the upper state total degeneracy, is given by the product of rotational and spin degeneracies: $g_{u} \equiv g_{J}g_{K}g_{I}$ \citep{2015mangum} \footnote{JPL gives different value of the nuclear spin degeneracy $g_{I}$ for HNCO, for the consistency, we use the value 1 for $g_{I}$ of HNCO.}. The values of $g_{u}$ for HNCO and H$^{13}$CO$^+$ are 9 and 3, respectively. Those values are from The Cologne Database for Molecular Spectroscopy (CDMS) \citep{2001muller,2005muller,2016endres}. $E_{l}$ is the energy of the lower level from Splatalogue \footnote{https://splatalogue.online}. $J_{\nu}(T_{ex})$ and $J_{\nu}(T_{bg})$ are the equivalent Rayleigh-Jeans excitation and background temperatures. 
  \begin{eqnarray}
J_\nu\left(T\right)&\equiv&\frac{\frac{h\nu}{k}}{exp\left(\frac{h\nu}{kT}\right)-1}.
\end{eqnarray}
$Q_{rot}$ is the partition function. The values of $Q_{rot}$ at two possible excitation temperatures 37.5\,K and 150\,K are listed with other parameters in Table \ref{tab:calculation}. The assumed excitation temperature 37.5\,K is consistent with NH$_{3}$ measurements towards the hot cores \citep[e.g.][]{2022kohno,2019billington,2019urquhart,2014hernandez,2012wienen}. The higher excitation temperature 150\,K is approximately the average temperature derived towards several hot cores in this observation \citep[e.g.][]{2008hunter,2011qiu,2019navarete,2020nguyenluong,2022beuther}. The column density ratios for $N[HNCO]/N[HCO^+]$ converting from $[HNCO]/[HCO^+]$ are 30.05 and 161.53 for the outflow line wings at excitation temperatures of 37.5\,K and 150\,K, respectively. 

In the line center where HCO$^+$ becomes optically thick, the column density of HCO$^+$ is converted from that of H$^{13}$CO$^+$ by a factor of 50. This value is within the range of 30 to 70 dependent on the distance and the environment in the Galaxy \citep{1994wilson} and similar to the conversion factor of 50 which has been used in previous observations in \citet{2006purcell,2012sanhueza,2011vasyunina}. An average of 55.6 for $^{12}$C/$^{13}$C  in all sources is derived using the most recent (4.77$\pm$0.81)R$_{GC}$+(20.76$\pm$4.61) in \citet{2023yan} with R$_{GC}$ from Table \ref{tab:parameters}. The column density ratios for $N[HNCO]/N[HCO^+]$ converting from $[HNCO]/[HCO^+]$ are thus 0.57 and 3.06 for the line central emission at excitation temperatures of 37.5\,K and 150\,K, respectively. The column density ratios for excitation temperature at 37.5\,K range from (7.92$\pm$0.30)\% to (44.11$\pm$1.76)\% with the median of (15.11$\pm$1.42)\%. Table \ref{tab:nratio} only lists the results for 37.5\,K considering the beam effects. The column density ratios of $N[HNCO]/N[HCO^+]$ at 150\,K can be scaled from those at 37.5\,K. The derived column density ratios are consistent with previous observations towards massive star-forming regions \citep{2011vasyunina,2012sanhueza}, where an average of $\sim$11\% were derived. Future observations of other transitions with higher resolution will help better constrain the excitation conditions which will give better view.

\subsection{Comments on Individual Sources}\label{subsec:disc2}

The associations of the HNCO (4--3) and HCO$^+$ (1--0) red- and/or blue-lobes distributions suggest that outflows from star-forming activities driven by the protostellar objects, either newly-formed or deeply embedded in H\uppercase\expandafter{\romannumeral2}/UCH\uppercase\expandafter{\romannumeral2} regions, could be the plausible explanation. The possible impacts from H\uppercase\expandafter{\romannumeral2} regions on the photodestruction of HNCO need further investigations. In this study, the spatial distributions of red- and/or blue-lobes of HNCO (4--3) are discussed with those of the high velocity tracer HCO$^+$ (1--0) and previous studies to determine the possible origins of the red and/or blue line wings. A common feature for all sources is the association in the spatial distribution of the red- and/or blue-lobes of HNCO (4--3) and HCO$^+$ (1--0), which might indicate the same origin of line wing broadening.

 {\it G005.88-0039} drives one of the most luminous bipolar outflows known in the Galaxy, which have been identified previously in tracers such as $^{13}$CO (2--1), SiO(5--4), HCO$^{+}$ (1--0), CO(2--1), and CO (3--2) \citep[e.g.][]{2004sollins,2007watson,2008hunter,2012su}. In this IRAM 30\,m observation, HNCO (4--3) shows collimated outflows with the same direction as in HCO$^+$ (1--0).

{\it G011.91-00.61} is located in an Infrared Dark Clouds (IRDCs), which have been considered to be representative of the earliest stages of massive star formation \citep[e.g.][]{2006simon,2009peretto,2021xie}. The central core of G011.91-00.61 is the brightest millimeter core in this IRDC \citep{2011cyganowski}. Bipolar outflows in $^{12}$CO (2--1), HCO$^{+}$ (1--0), and SiO (2--1) have been observed towards the central core \citep{2011cyganowski}, which are consistent with the well-collimated outflows traced by HNCO (4--3) and HCO$^+$ (1--0) in this observation.

{\it G012.80-00.20} resides within the W33 complex, first detected as a thermal radio source in the 1.4\,GHz \citep{1958westerhout}. The molecular gas in the expanding shell around H\uppercase\expandafter{\romannumeral2}  region has been suggested to be compressed, heated, and driven by the expansion of the H\uppercase\expandafter{\romannumeral2}  region \citep[e.g.][]{1989keto,2022khan}. Several sources were identified with high angular resolution continuum mappings with SMA \citep{2014immer}. 
 Among those sources, the northern fainter one AGAL012.804-00.199, was suggested to drive an outflow from CO observations \citep{2019navarete}. With the same central velocity of CO, the outflows seen in HNCO (4--3) and HCO$^+$ (1--0) in this observation could also be originated from AGAL012.804-00.199.

{\it G015.03-00.67} is located at the edge of the ionization front of M17 \citep{1980chini}. Blueshifted emissions of HNCO (4--3) and HCO$^+$ (1--0) are dominant in the region, which could be influenced by the Young Stellar Objects (YSOs) identified from {\it Spitzer} \citep{2009povich} and {\it Chandra} \citep{2007broos} and/or the OB stars in the open cluster NGC 6618 towards the H\uppercase\expandafter{\romannumeral2}  regions \citep{2022bordier,2008hoffmeister,1997hanson,1980chini}. A likely driving source is AGAL015.029-00.669, which is associated with water and Class\uppercase\expandafter{\romannumeral1} methanol masers and revealed to show outflow in CO \citep{2019navarete}.

 {\it G034.49+00.22} is another source located in an IRDC, in addition to G011.91-00.61. Hot cores/UCH\uppercase\expandafter{\romannumeral2}  regions and a B0.5 protostar have been identified towards this region \citep{2005rathborne,1973panagia}. Both HNCO (4--3) and HCO$^+$ (1--0) display bipolar outflow features towards the southern core. The excess 4.5 $\mu$m emission towards the MM4 core reported by \citet{2005rathborne} was considered to be produced by ionized gas and/or shocked gas \citep{2009chambers}. The two dense regions depicted in HC$_{3}$N (10--9) emissions correspond to MM2 and MM4 in \citet{2005rathborne,2006rathborne}, respectively. MM2 is associated with IRAS 18507+0121 \citep{2004shepherd}, together with water and methanol masers \citep{1994miralles}. Spatially overlapping blueshifted and redshifted emissions were observed towards MM2 and MM4 \citep{2007shepherd,2010sanhueza}. Self-absorption emissions can be seen on HCO$^+$ in Figure~\ref{fig:spec_hcop}, which is consistent with CO (3--2) and HCO$^+$ spectra in \citet{2010sanhueza}. Shocks in MM2 and MM4 can also be supported by the SiO detections shown in Figure~\ref{fig:outflow}.

{\it G075.76+00.33} is located at the southern rim of the H\uppercase\expandafter{\romannumeral2} region G075.78+0.34 coinciding with a compact core seen in HCN \citep{2010riffel}, which could be associated with a YSO seen by \citet{2013cooper}. The directions of the blue-shifted emissions of HNCO and HCO$^+$ are towards the southwest densest region.

{\it G109.87+02.11} is an active massive star-forming region detected with several YSOs \citep{1984hughes,1994rodriguez}. The bright radio source HW2, is associated with an early-B type star \citep{1984hughes}. Intense magnetic fields and very bright masers were inferred to be associated with HW2 \citep[e.g.][]{2005bartkiewicz,2006vlemmings,2010vlemmings}. This source is the second detected RRL maser to date \citep{2011jimenezserra}. A collimated, high-velocity ionized jet and pulsating outflows have been detected in $^{12}$CO(2--1), $^{12}$CO(1--0), and SiO(2--1) \citep{2007comito,2009cunningham,2013zapata}. Though HCO$^+$ spectrum shows both blue- and red-shifted emission and a collimated bipolar outflow, HNCO only shows blueshifted emission.

{\it G111.54+00.77} is related with one of the radio continuum source IRS 1 located in an optically visible H\uppercase\expandafter{\romannumeral2}  region NGC7538 \citep{1974wynnwilliams,1984campbell,2009moscadelli}. The central star of IRS 1 was suggested to be an O7 or earlier \citep{1984campbell, 2010puga,2020sandell}. An inverse P-Cygni profile indicative of infall was seen in HNCO (10$_{0,10}$--9$_{0,9}$) \citep{2011qiu}. While still accreting mass, this very young star is considered to be the driving source of an ionized jet and the hypercompact H\uppercase\expandafter{\romannumeral2}  region \citep{2020sandell}. The distribution of HNCO (4--3) shows a similar shape to the previously identified bubble-shape in C$^{17}$O (3--2), H$^{13}$CO$^{+}$ (4--3), continuum ($\sim$340\,GHz), and CII emission \citep{2011qiu,2014frau,2022beuther}. These morphology features could be resulted from protostellar feedback \citep{2014frau} and/or the strong FUV radiation \citep{2020sandell}. The FUV radiation indicated by CII emission from PDR could destroy HNCO that shapes HNCO emission into a bubble. The spatial distributions of HNCO blue lobes show different morphologies from those of HCO$^+$ blue lobes, which could be most likely due to the low signal to noise ratio. The intrinsic differences in astrochemical properties could also contribute to the morphological differences between the HNCO and HCO$^+$ blue lobes.

{\it G192.60-00.04} is within Sh 2-255, the optically bright H\uppercase\expandafter{\romannumeral2}  region first cataloged by \citet{1959sharpless}. The northern region of the mapping in this observation is adjacent to S255 N, which hosts a UCH\uppercase\expandafter{\romannumeral2}  region and is associated with methanol and water masers \citep{2012zinchenko,2004kurtz,2007cyganowski,2018zemlyanukha}. S255 N was considered to drive a collimated outflow \citep{1997miralles,2007cyganowski}, while global collapse was observed in HCO$^+$ (3--2) \citep{2007minier}. The center of the northern region corresponds to S255N-SMA6 in \citet{2012zinchenko}, where broad SiO emission feature and CO outflow were detected \citep{2012zinchenko}. The southern region is related to S255 IR, a hot core associated with methanol and water masers \citep{2007goddi,2010rygl}, as well as several compact H\uppercase\expandafter{\romannumeral2}  regions \citep{1986snell}. Disc-mediated accretion has been proposed to explain the multiple burst events observed in radio to optical bands towards YSO in this region \citep[e.g.][]{2015fujisawa,2017carattiogaratti,2018liushengyuan}. Higher transitions of HNCO (10--9) were detected towards S255 IR \citep{2011wang,2015zinchenko}, where an abundance of $\sim$10$^{-8}$ was derived for HNCO \citep{2015zinchenko}. The northeast to southwest collimated bipolar outflows seen in CO and H$_{2}$ \citep{2011wang} are consistent with the HCO$^+$ outflow direction in this observation.

\section{Discussion}\label{sec:discussion}

\subsection{Velocity ranges of the line wings}\label{subsec:disc1}

The velocity ranges of HNCO (4--3) line wings appear smaller than that of other shock tracers \citep[e.g.][]{2000zinchenko,2010rodriguezfernandez,2013sanchezmonge}. The line wings of HCO$^+$ (1--0) listed in Table \ref{tab:range} are larger than those of HNCO (4--3) by 7\,km s$^{-1}$ on average. In three sources, red- or blue-shifted emissions are seen in HCO$^+$ (1--0), but not in HNCO (4--3). The critical density of HCO$^+$ (10$^{4}$--10$^{5}$\,cm$^{-3}$) \citep{2015shirley} is about one order of magnitude lower than that of HNCO (10$^{5}$--10$^{6}$\,cm$^{-3}$) \citep{1991nguyenqrieu}, which could explain the excess of red and/or blue line wing in HCO$^{+}$ (1--0).

The line wings of HNCO (4--3), as well as HCO$^+$ (1--0), are determined from comparisons on the intensities with optically thin tracers HC$_{3}$N (10--9) and H$^{13}$CO$^+$ (1--0). However, HC$_{3}$N (10--9) or H$^{13}$CO$^+$ (1--0) might not always be optically thin and unaffected by outflows. Red- and blue-line wings can be distinguished on HC$_{3}$N (10--9) profiles towards G011.91-00.61. A similar feature has been observed towards low-mass star-forming region L1157 \citep{2004beltran} where HC$_{3}$N abundance got enhanced by a factor of 25 and 50 \citep{1997bachiller,2018mendoza}. The factor of HNCO abundance enhancement towards the same source was derived to be 6-83 \citep{2010rodriguezfernandez}. Chemical modeling which includes shock code supported HC$_{3}$N could be produced by gas phase reactions when shock passes \citep{2018mendoza}. For another optically thin tracer H$^{13}$CO$^+$ (1--0), the clear red-shifted line wing towards G005.89-00.39 and a blue-shifted peak with a red-shifted shoulder as described in \citet{2005devries} could be identified towards G109.87+02.11. The possible line wings of HC$_{3}$N (10--9) and H$^{13}$CO$^+$ (1--0) might result in an underestimate of the line wing ranges of HNCO (4--3) and HCO$^+$ (1--0).

   \subsection{Formation mechanism in shocks}\label{subsec:formation}

The formation mechanism of HNCO, a peptide link (-NH-C(=O)-) molecular specie essential to living systems, has drawn attention to chemical modelings, including gas-phase reactions \citep[e.g.][]{1977iglesias}, grain surfaces models \citep[e.g.][]{2010quan}, and those involving shocks \citep[e.g.][]{2020zhangxia}. The observed line wings and the substantial intensities in the line wings of HNCO in these sources are most likely resulted from shocks, which play the role in both sputtering HNCO from dust grains and forming HNCO in the gas-phase \citep{2010rodriguezfernandez,2016burkhardt}. Prior to heating or the arrival of shocks, the relatively long timescale allows a variety of chemical reactions to form HNCO on grain surfaces even at low temperatures. It has been recognized that the major formation mechanism for HNCO on dust grains is the isocyanate radical NCO accreted onto the grain to react with the surface hydrogen, which is relatively free compared to larger molecular species at low temperatures  \citep{2008garrod,2010quan}. Another formation mechanism involving thermal reaction NH+CO$\ce{->}$HNCO is also plausible in hot cores \citep{2008garrod,2010tideswell}. Shocks, even at low velocity, can efficiently release the abundant HNCO from dust grains into gas-phase \citep{2010quan,2020zhangxia}. The temperature in the shocked gas then enables the efficient hydrogenation of NCO \citep{2000zinchenko,2010rodriguezfernandez}. The abundance of HNCO is also considered to be related to shock age and the grain mantle composition at the time of shock arrival \citep{2010rodriguezfernandez,2020zhangxia}. High intensity ratios of HNCO to HCO$^+$ in the line wings could be explained from the recent shock chemical modeling results that the abundance of HNCO can be increased even in weakly shocked regions \citep{2020zhangxia}. Further detailed chemistry modelings are needed to fully understand HNCO in the shocks in these hot cores.


\section{Summary}\label{ref:summary}

We carried out mapping observations of HNCO, HCO$^+$, HC$_{3}$N, and H42$\alpha$ toward nine hot cores in a sample of twenty three sources from \citet{2014reid} with the IRAM 30 m telescope. The main results are:
\begin{enumerate}
	\item The line wing emissions of HNCO are imaged for the first time. We found that the spatial distributions of the red- and/or blue-lobes of HNCO emission nicely associate with those lobes of HCO$^+$ emission, indicating that HNCO is a good tracer of outflow.
	\item High intensity ratios of HNCO (4--3) to HCO$^+$ (1--0) are derived in the line wings, suggesting that HNCO could efficiently form in the shock environment.
	\item The derived column density ratios of HNCO to HCO$^+$ in the line center, ranging from (7.92$\pm$0.38)\% to (44.11$\pm$1.76)\%, are consistent with those previously observed towards high-mass star-forming regions.
\end{enumerate}

\begin{acknowledgments}
We would like to thank the anonymous referee for his/her critical comments and constructive suggestions, which has significantly improved this work. This work is supported by the national Key R\&D Program of China (No. 2022YFA1603101), the National Natural Science Foundation of China (NSFC, Grant No. 11988101 and No. U1731237). This study is based on observations carried out under project number 042-19, 147-19, and 127-20 with the IRAM 30-m telescope. IRAM is supported by INSU/CNRS (France), MPG (Germany), and IGN (Spain). 
\end{acknowledgments}

%

\vspace{5mm}
\facilities{IRAM}


\software{class\citep{2005pety}
          }

\clearpage
\begin{sidewaystable}
\centering
\caption{Source information and observing parameters}\label{tab:parameters}
\vspace{-1mm}
\begin{tabular}{lcccccccl} 
\hline
Source & Alias & R.A. & Decl. & Region Type &   D (kpc) & $R_{GC}$ (kpc)   & Mapping Size & Reference\\
\hline
G05.88-00.39 & W28 & 18:00:30.31 & -24:04:04.5 & UCH\uppercase\expandafter{\romannumeral2}  & 3.0   & 5.3 & 80$''\times$90$''$ &  \citet{2014sato}\\
G011.91-00.61 & IRDC G11.92-0.61 &  18:13:59.72& -18:53:50.3  &  HMPO & 3.4 &  5.1&  90$''\times$130$''$ &  \citet{2014sato}\\
G012.80-00.20 & W33 Main& 18:14:14.23 & -17:55:40.5 &UCH\uppercase\expandafter{\romannumeral2}  & 2.9  &  5.5 &  100$''\times$100$''$  &  \citet{2013immer} \\
G015.03-00.67 & M17 & 18:20:22.01 & -16:12:11.3 & HCH\uppercase\expandafter{\romannumeral2} & 2.0  & 6.4 & 100$''\times$100$''$  &  \citet{2011xu}\\
G034.39+00.22 & IRDC G34.43+0.24 MM4 & 18:53:19.00 & +01:24:50.8 & UCH\uppercase\expandafter{\romannumeral2} &   1.6  & 7.1   & 90$''\times$100$''$ & \citet{2011kurayama}\\
G075.76+00.33 & Onsala 2&20:21:41.09& +37:25:29.3 & H\uppercase\expandafter{\romannumeral2}  &  3.5     & 8.2 & 100$''\times$100$''$ &   \citet{2013xu} \\
G109.87+02.11 & CepA HW 2&22:56:18.10& +62:01:49.5 & UCH\uppercase\expandafter{\romannumeral2} &   0.7  & 8.6 &100$''\times$100$''$ &  \citet{2009moscadelli} \\
G111.54+00.77 & NGC7538 IRS1& 23:13:45.36 & +61:28:10.6 & UCH\uppercase\expandafter{\romannumeral2} &   2.6 & 9.6  &  100$''\times$100$''$ & \citet{2009moscadelli} \\
G192.60-00.04 & S255 & 06:12:54.02 & +17:59:23.3 & UCH\uppercase\expandafter{\romannumeral2} &  1.6    & 9.9   &  100$''\times$100$''$  &  \citet{2010rygl} \\
\hline
\end{tabular}
\end{sidewaystable}
\clearpage

\begin{table}
\centering
\caption{Physical parameters for HNCO, HCO$^+$, H$^{13}$CO$^+$, and HC$_{3}$N}\label{tab:calculation}
\begin{tabular}{lcccccr} 
\hline
Molecular Line & $\lambda$ & A & $B_{0}$ & $E_{l}/k$ & $Q_{rot}$ & $Q_{rot}$ \\
 & cm & 10$^{-5} s^{-1}$ & MHz & K  & 37.5\,K & 150\,K\\
\hline
HNCO (4--3)& 0.340963 & 0.878011  & 11071.7 &   6.32914  & 117.3039 & 2820.206\\
HCO$^+$ (1--0) &  0.336133 & 4.18697  & 44594.4 & 0 & 17.8601 & 70.4887  \\
H$^{13}$CO$^+$ (1--0)& 0.345565  &  3.85389  & 43377.3   &0 & 18.3516 & 72.4065 \\
HC$_{3}$N (10--9) & 90978.989 & 5.8125 & 4549.059 & 19.64782 & 172.1063 & 687.5086 \\
\hline
\end{tabular}
\end{table}

\clearpage 

\begin{sidewaystable}
\begin{center}
\scriptsize
\caption{Velocity ranges and integrated intensities for HNCO(4--3) and HCO$^{+}$(1--0)$^{a}$ towards the center and lobes}
\label{tab:range}
\setlength{\tabcolsep}{2pt}
 \begin{tabular}{lccccccccccc}
  \hline\noalign{\smallskip}
  \hline\noalign{\smallskip}
Source & \multicolumn{3}{c}{center}                                                                                         & \multicolumn{4}{c}{red}                                   & \multicolumn{4}{c}{blue}                                \\ 
   &     HNCO, HCO$^+$    &   $I_{\rm HNCO}$  &  $I_{\rm H^{13}CO^{+}}$      & HNCO(4--3)  &  $I_{\rm HNCO}$  & HCO$^+$(1--0)   &  $I_{\rm HCO^{+}}$   &HNCO(4--3)   &  $I_{\rm HNCO}$  & HCO$^+$(1--0)  & $I_{\rm HCO^{+}}$   \\
                      & km s$^{-1}$     &  K km s$^{-1}$  & K km s$^{-1}$   &  km s$^{-1}$   &   K km s$^{-1}$  &  km s$^{-1}$  & K km s$^{-1}$    & km s$^{-1}$ &  K km s$^{-1}$    &  km s$^{-1}$  &   K km s$^{-1}$ \\
    \hline\noalign{\smallskip}
G05.88-00.39            &    [5,15]   &  2.33$\pm$0.11 &  16.78$\pm$0.19  &   [15,27]&   1.22$\pm$0.12  &  [15,40]   &  12.22$\pm$0.45  &   [-17,5]   &  0.72$\pm$0.16  & [-20,5]  &    24.22$\pm$0.43    \\             
G011.91-00.61 &   [30,42]  &   3.80$\pm$0.12   &       4.91$\pm$0.12    & [42,55] &   1.34$\pm$0.13  & [42,55] &    5.74$\pm$0.13    & [18,30] &    1.03$\pm$0.12     &  [18,30]  &   5.36$\pm$0.12  \\
G012.80-00.20  &    [32,41] & 3.20$\pm$0.29  &    11.19$\pm$0.23  & [41,45] &   0.70$\pm$0.19   &  [41,49] &   8.67$\pm$0.28  & [28,32] &  0.86$\pm$0.19   & [28,32]  &    8.03$\pm$0.20   \\
G015.03-00.67 & [16,24]  &    2.00$\pm$0.12  &    10.73$\pm$0.13  & - & - & [24,30]&    3.50$\pm$0.39 & [0,16] &   0.80$\pm$0.16  & [10,16] &   6.53$\pm$0.39    \\
G034.39+00.22 MM2& [53,63]  &    1.22$\pm$0.11 &    5.00$\pm$0.14  & [63,70] &    0.49$\pm$0.10    & [63,80]    &    5.23$\pm$0.21   & - & - & -& - \\
G034.39+00.22 MM4 & [53,64] &   3.27$\pm$0.13 &   8.09$\pm$0.13 &  [64,68] &      0.49$\pm$0.08   &  [64,75] &   3.89$\pm$0.14  & [51,53]  &    0.27$\pm$0.05   & [50,53] &    0.74$\pm$0.08   \\
G075.76+00.33 &  [-6,3] &    0.87$\pm$0.20 &   4.32$\pm$0.20    & - & - & - & -& [-24,-6] &     0.93$\pm$0.28    & [-15,-6] &    7.82$\pm$0.22  \\
G109.87+02.11 & [-15,-8]&    1.10$\pm$0.15   &    7.85$\pm$0.17   & -& - & [-8,5] &   28.52$\pm$0.25  & [-23,-15] &  0.85$\pm$0.16   &[-40,-15] &  14.97$\pm$0.35   \\
G111.54+00.77 & [-63,-53] & 1.10$\pm$0.12 & 2.82$\pm$0.20  & - &- &  [-53,-47]  & 8.03$\pm$0.19   & [-75,-63] &  0.67$\pm$0.19 & [-73,-62] &  6.05$\pm$0.23 \\
G192.60-00.04 & [5,10] &0.59$\pm$0.09   &  1.28$\pm$0.10  & [10,16] & 0.39$\pm$0.12   & [10,13] & 2.81$\pm$0.09  & [1,5] & 0.08$\pm$0.08 & [1,5] &  2.55$\pm$0.10 \\
  \noalign{\smallskip}\hline
\end{tabular}
\end{center}
Notes: $^{a}$ The integrated intensities of HCO$^+$ (1--0) towards the center are derived from $I_{\rm H^{13}CO^{+}}$ by multiplying a conversion factor of 50.
\end{sidewaystable}
\clearpage

\begin{table}
\begin{center}
\caption{Column Density Ratios $N[HNCO]/N[HCO^+]$ towards the line center at 37.5\,K}
\label{tab:nratio}
\setlength{\tabcolsep}{2pt}
 \begin{tabular}{lc}
  \hline\noalign{\smallskip}
  \hline\noalign{\smallskip}
Source & center                                                                                          \\ 

    \hline\noalign{\smallskip}         
G05.88-00.39          &  (7.92$\pm$0.38)\%     \\   
G011.91-00.61 &       (44.11$\pm$1.76)\%    \\
G012.80-00.20 &  (16.30$\pm$1.52)\%     \\
G015.03-00.67 &    (10.62$\pm$0.65)\%    \\
G034.39+00.22 MM2&   (13.91$\pm$1.31)\%     \\
G034.39+00.22 MM4 &  (23.04$\pm$0.99)\%      \\
G075.76+00.33 &   (11.47$\pm$2.69)\%   \\
G109.87+02.11 &   (7.99$\pm$1.10)\%     \\
G111.54+00.77 & (22.23$\pm$2.90)\%    \\
G192.60-00.04 & (26.27$\pm$4.50)\%   \\
  \noalign{\smallskip}\hline
\end{tabular}
\end{center}
\end{table}
\clearpage

\clearpage
\begin{figure*}
\gridline{\fig{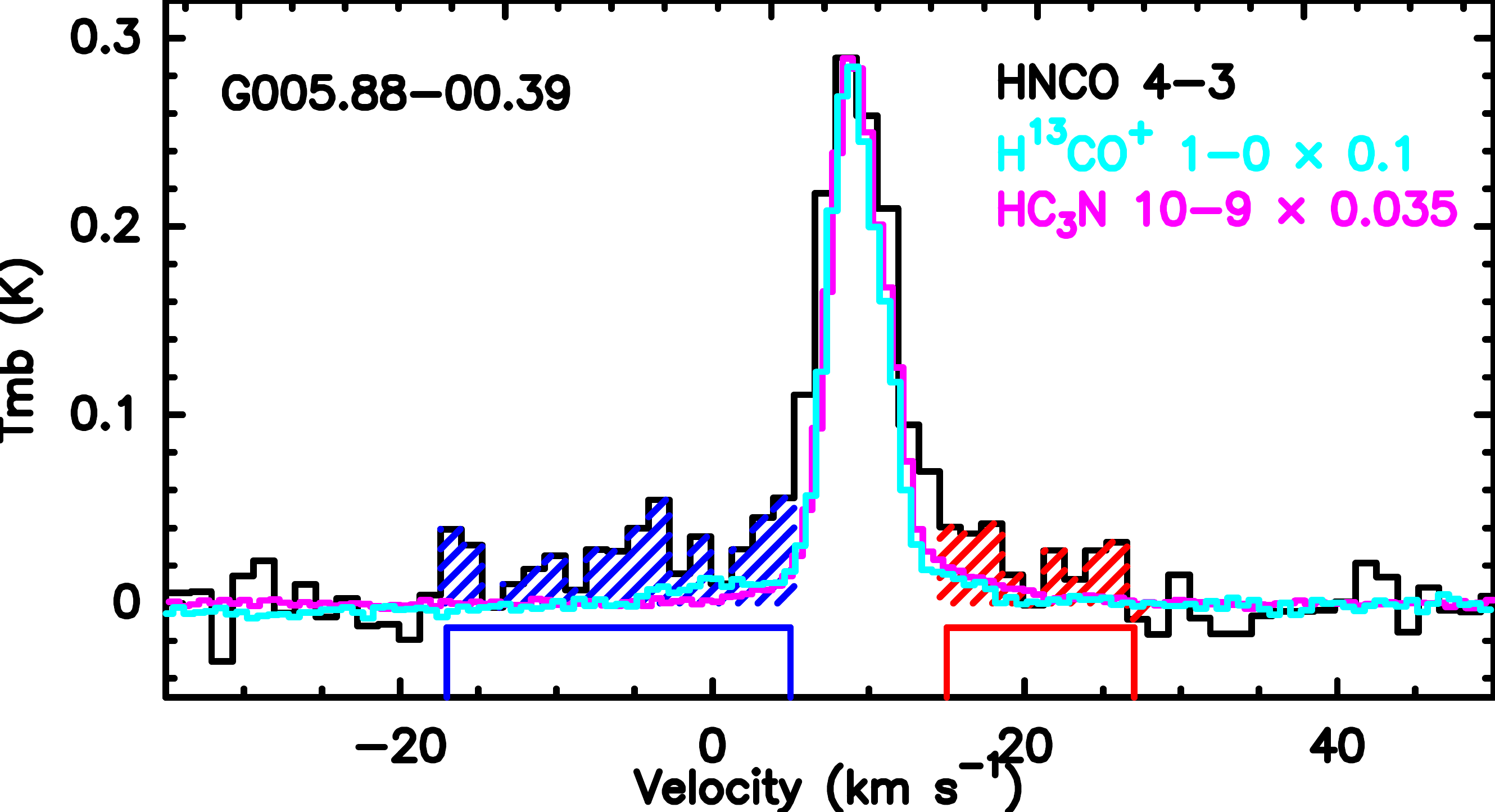}{0.38\textwidth}{(a)}
          \fig{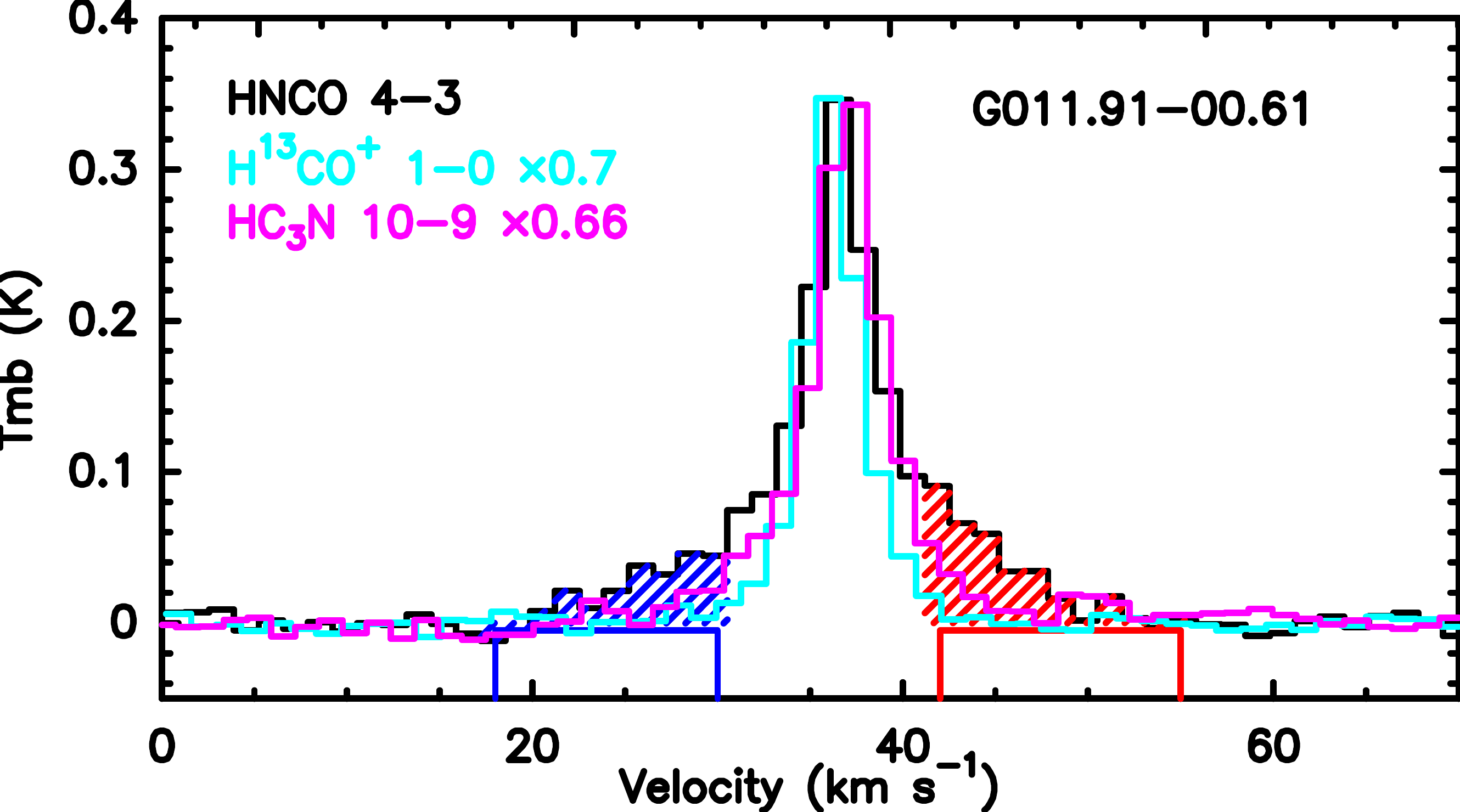}{0.38\textwidth}{(b)}}
\gridline{          
          \fig{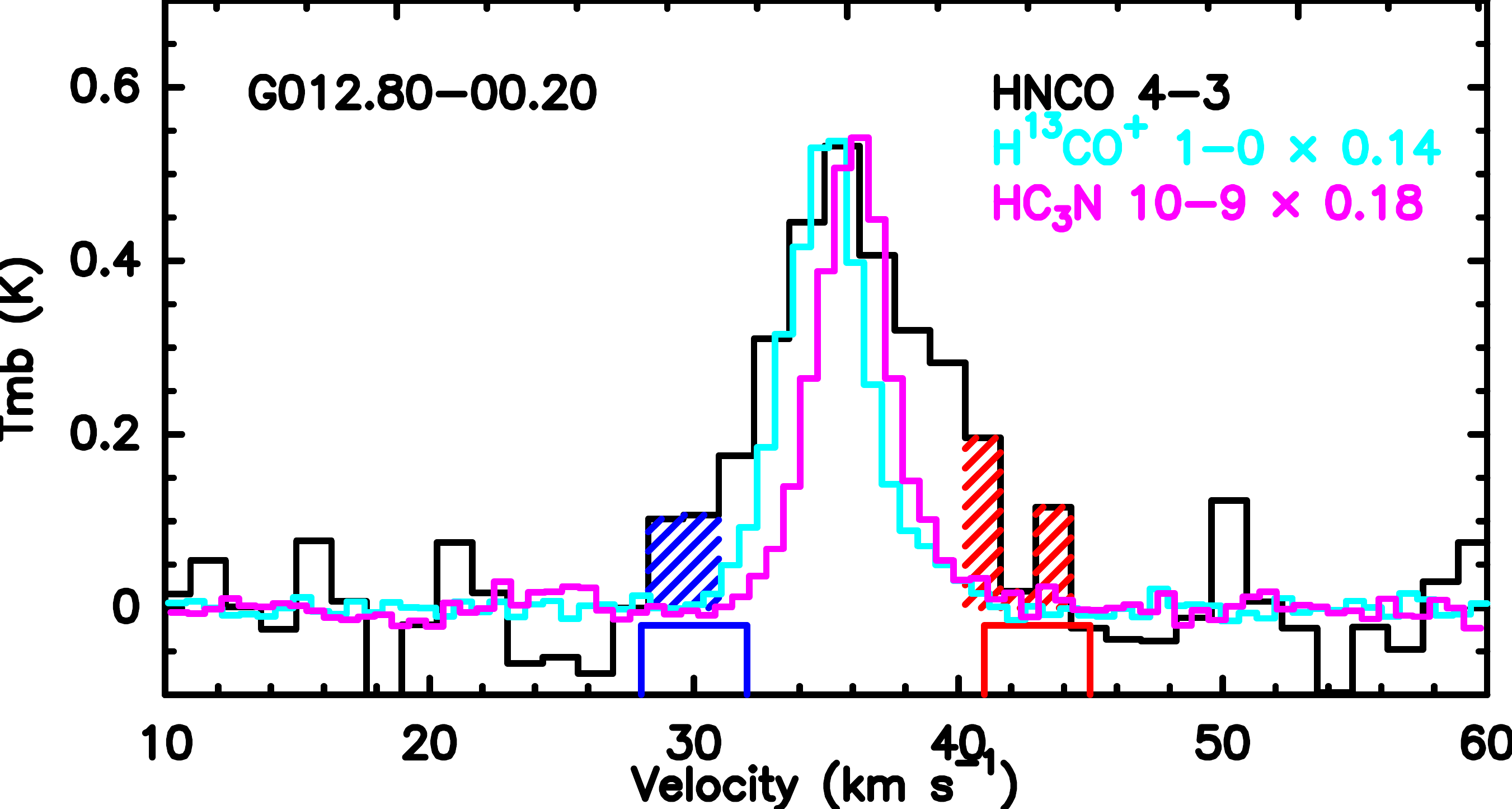}{0.38\textwidth}{(c)}
          \fig{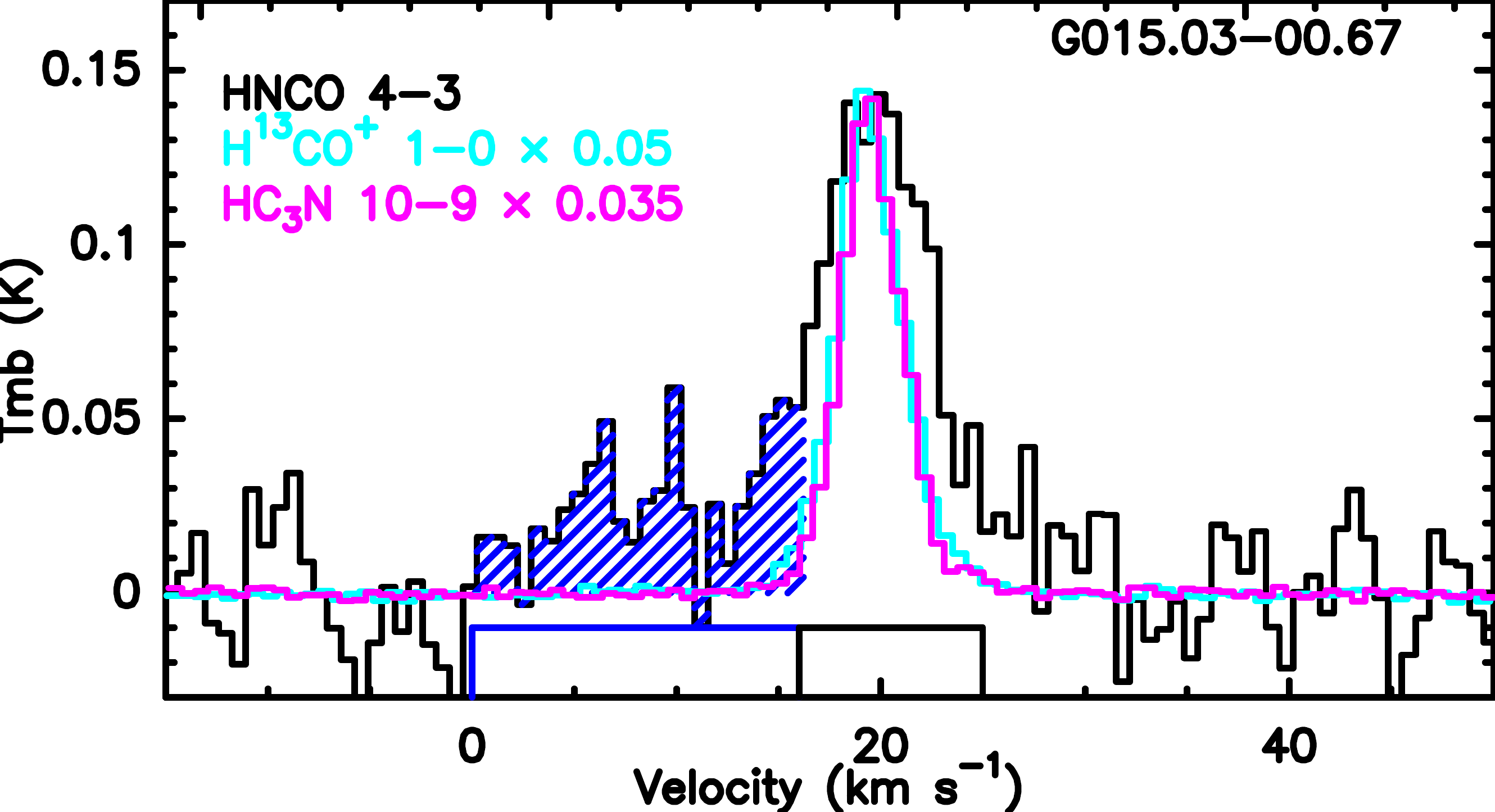}{0.38\textwidth}{(d)}
          }
\gridline{          
          \fig{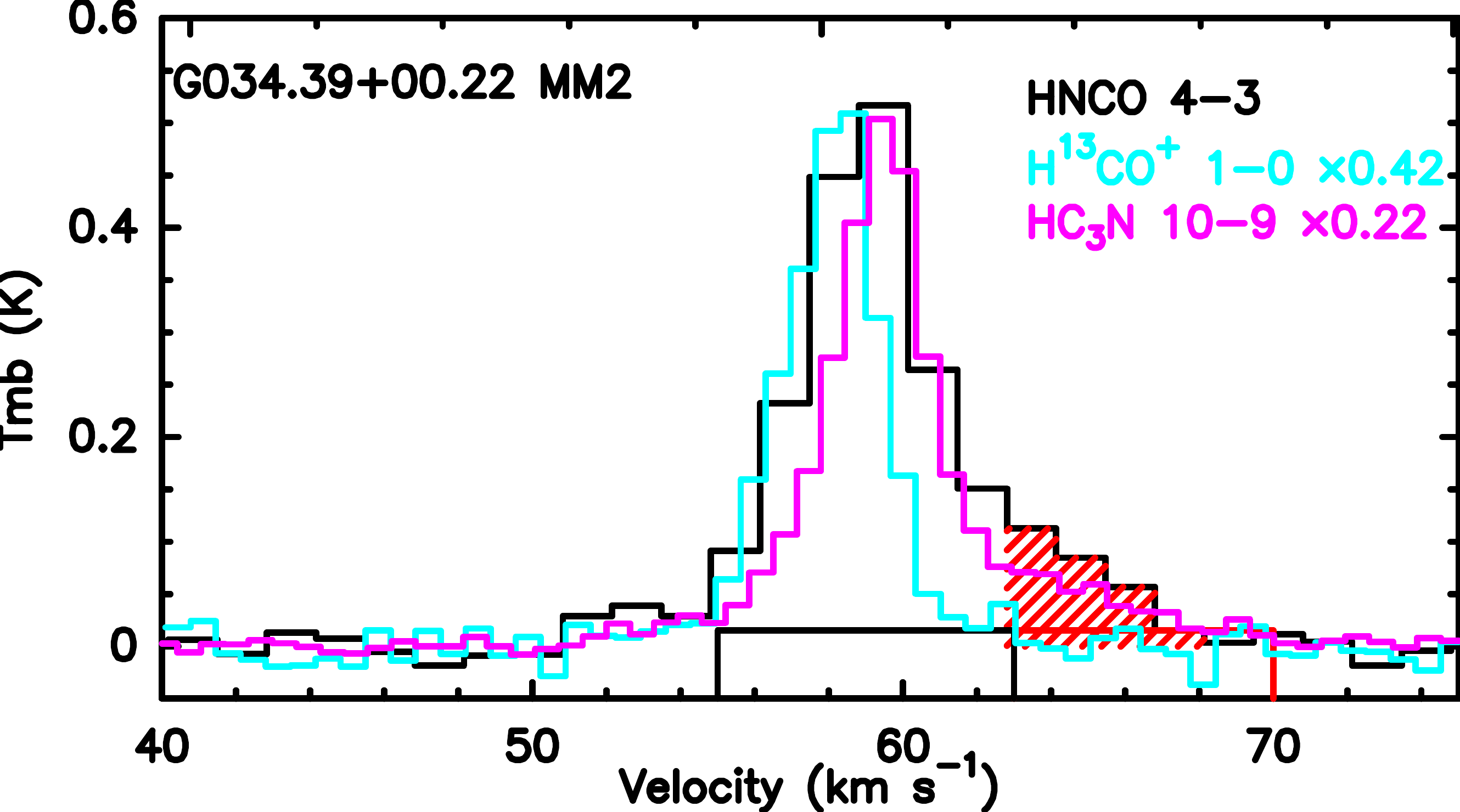}{0.38\textwidth}{(e)}
          \fig{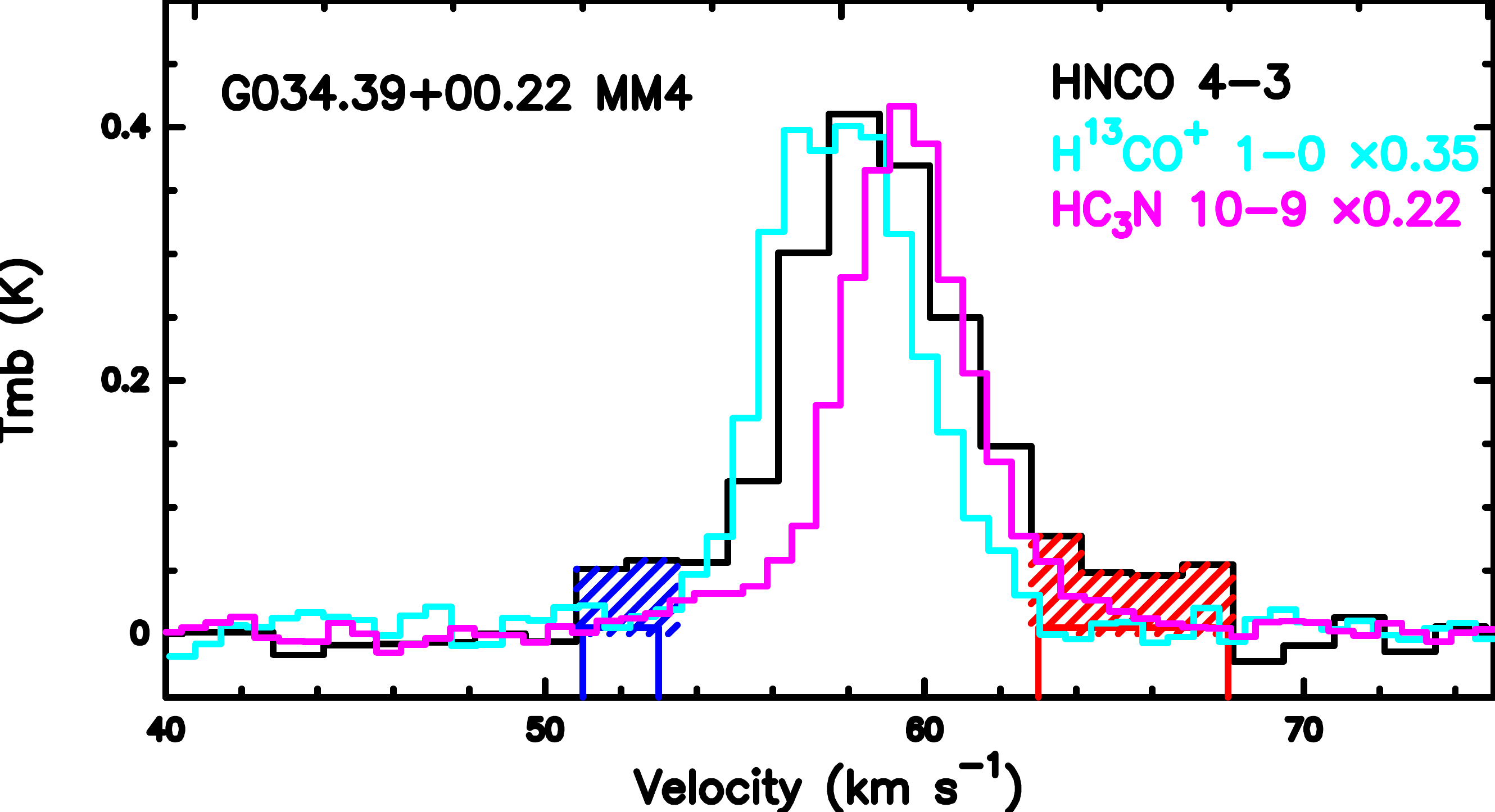}{0.38\textwidth}{(f)}
          }
\gridline{
          \fig{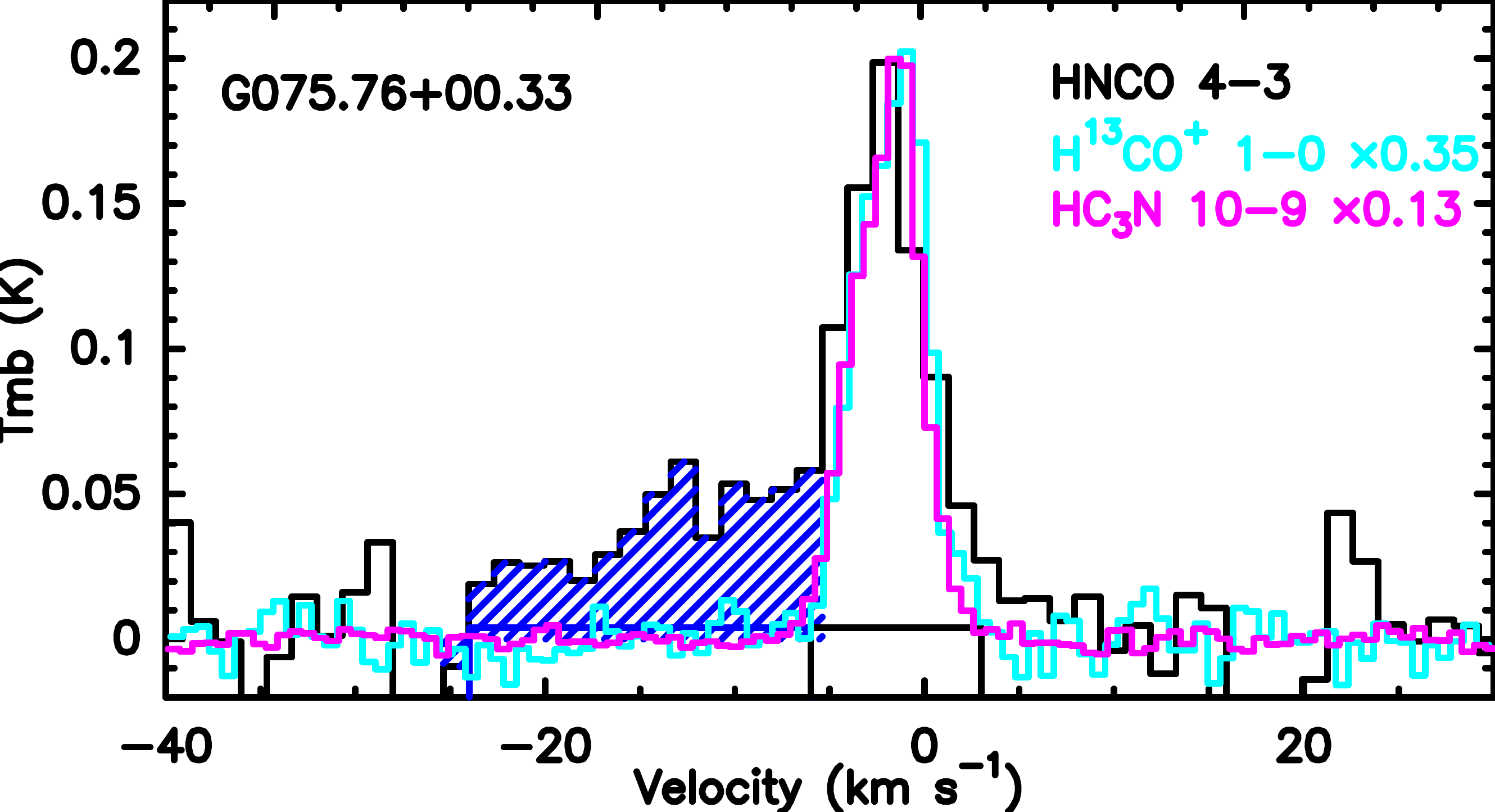}{0.38\textwidth}{(g)}
           \fig{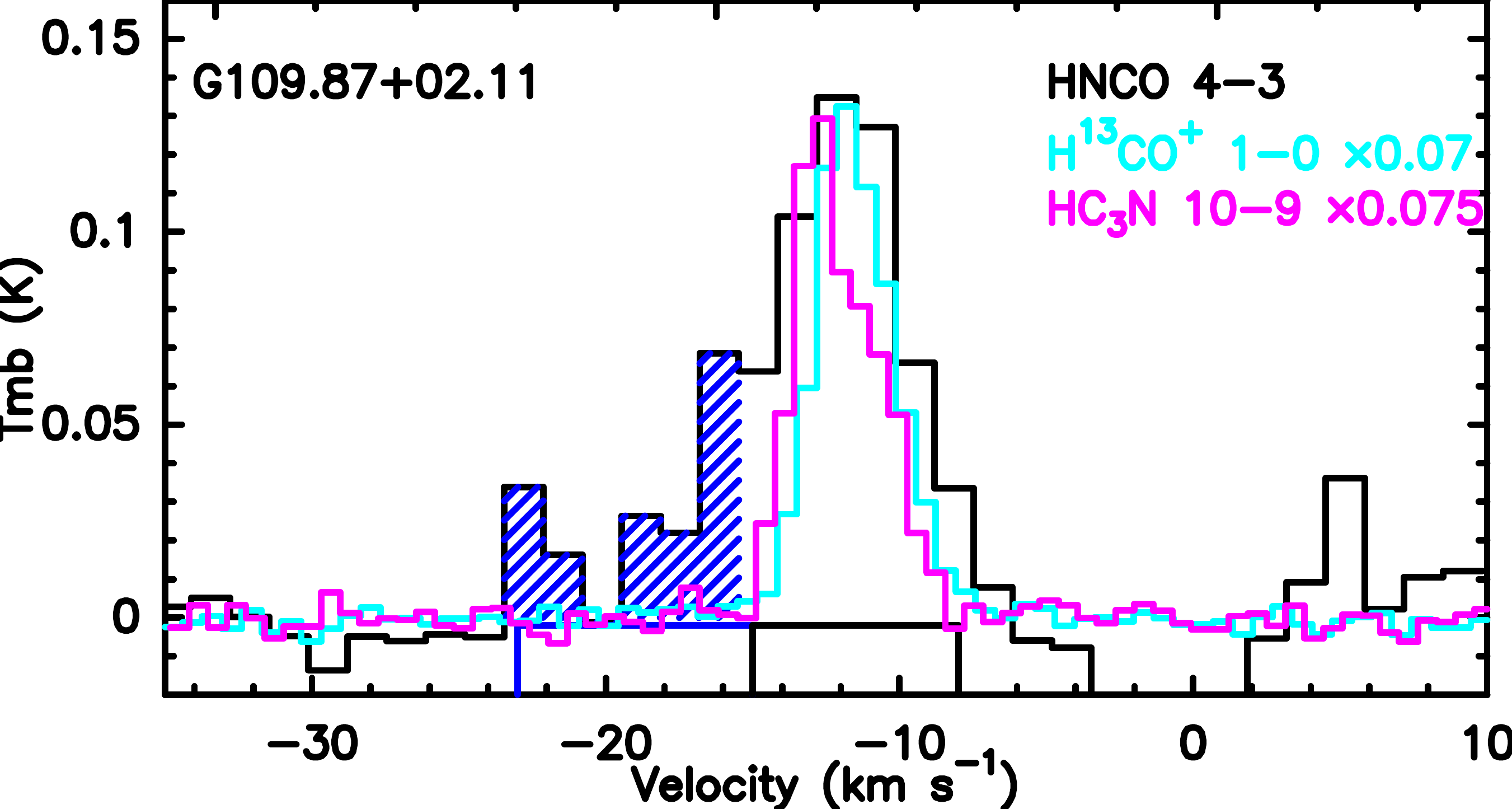}{0.38\textwidth}{(h)}
           }
 \gridline{\fig{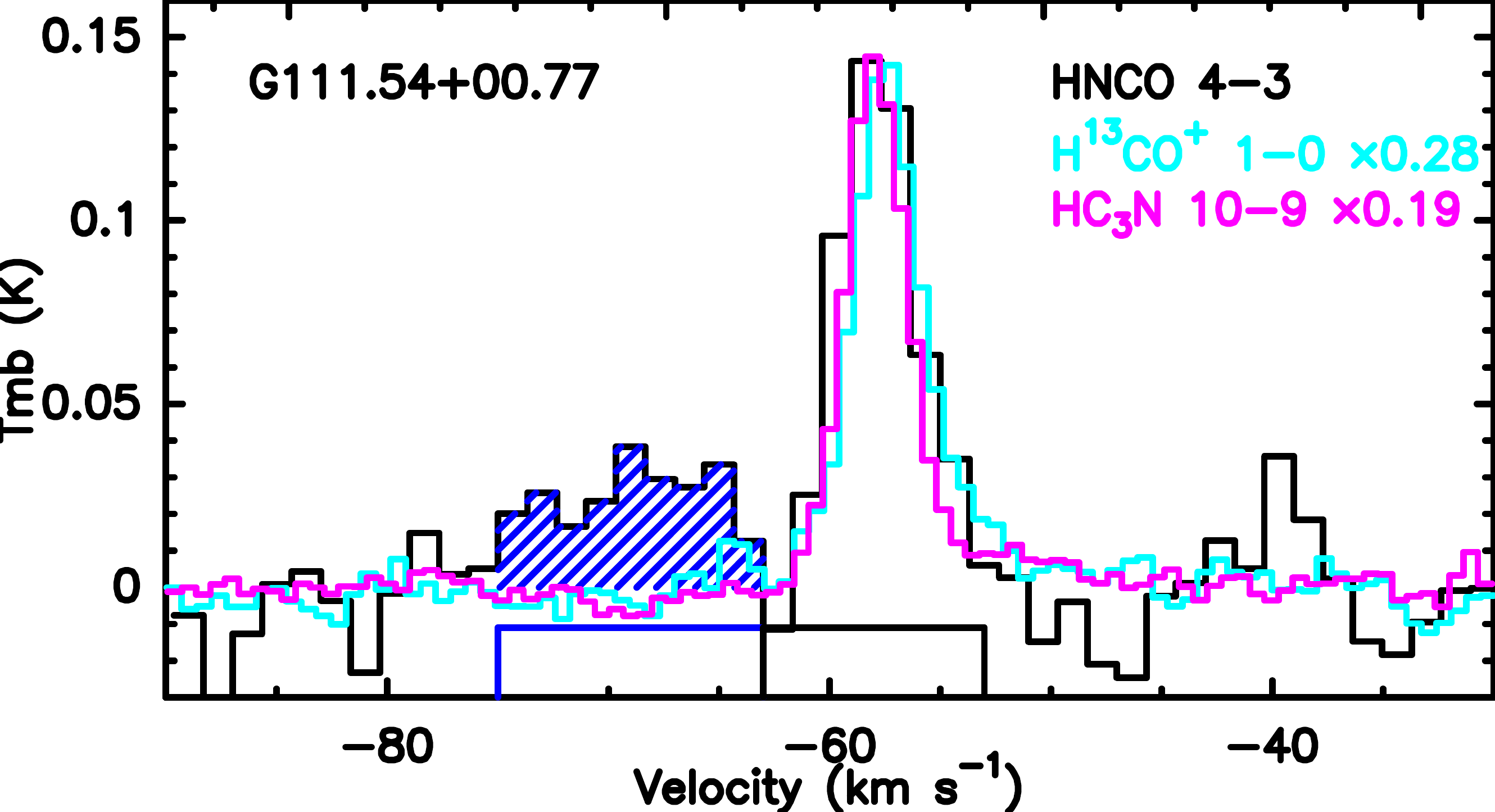}{0.38\textwidth}{(i)}
          \fig{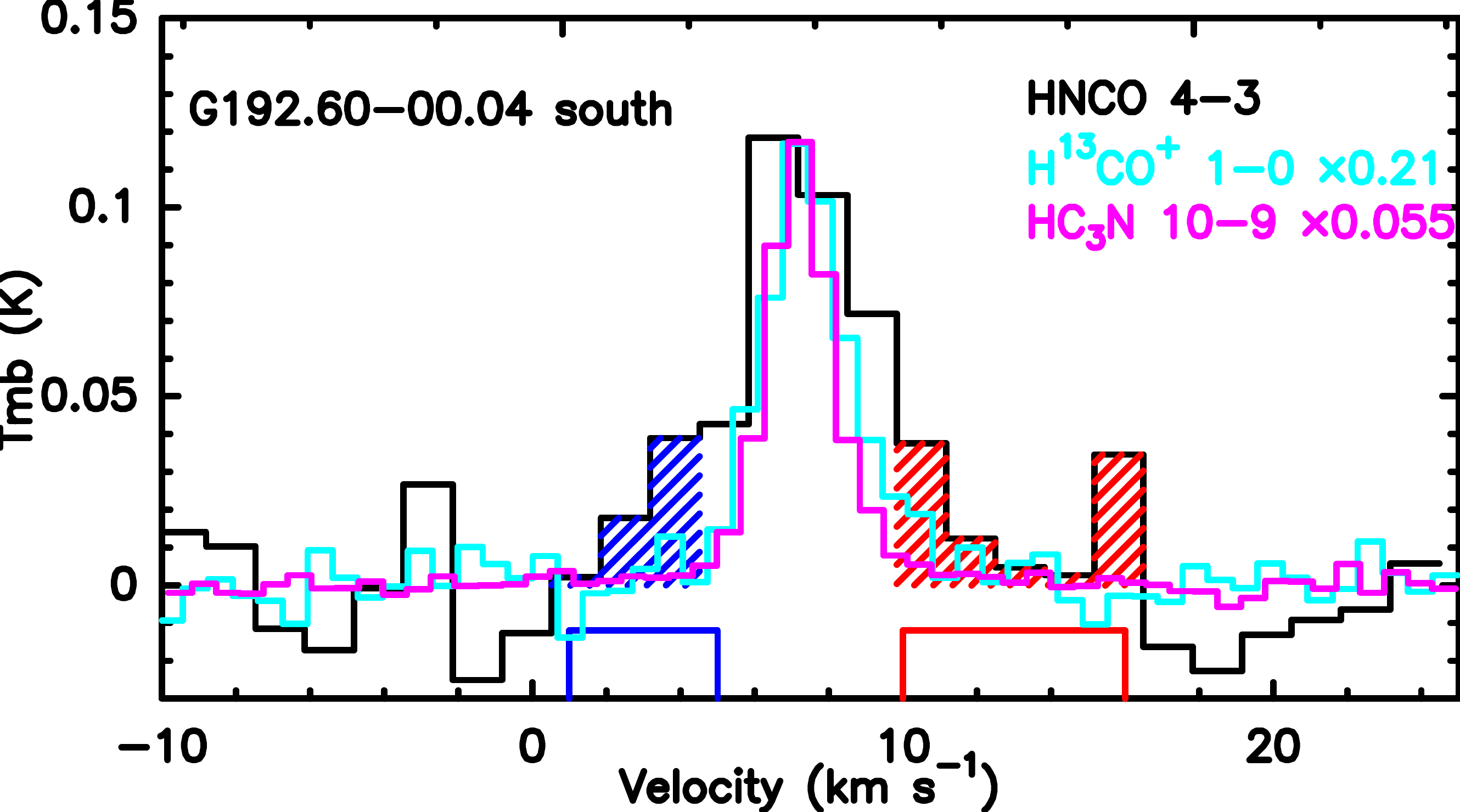}{0.38\textwidth}{(j)}
               }
\caption{The spectra of HC$_{3}$N (10--9) and H$^{13}$CO$^{+}$ (1--0) normalized to the peak intensity of HNCO (4--3) towards each source. The blue and red line wing emissions are labelled with blue and red windows. The black windows refer to the line core emission ranges. MM2 and MM4 in (e) and (f) refer to the two dust cores identified in \citet{2005rathborne}.}
\label{fig:spec}
\end{figure*}

\clearpage
\begin{figure*}
\gridline{\fig{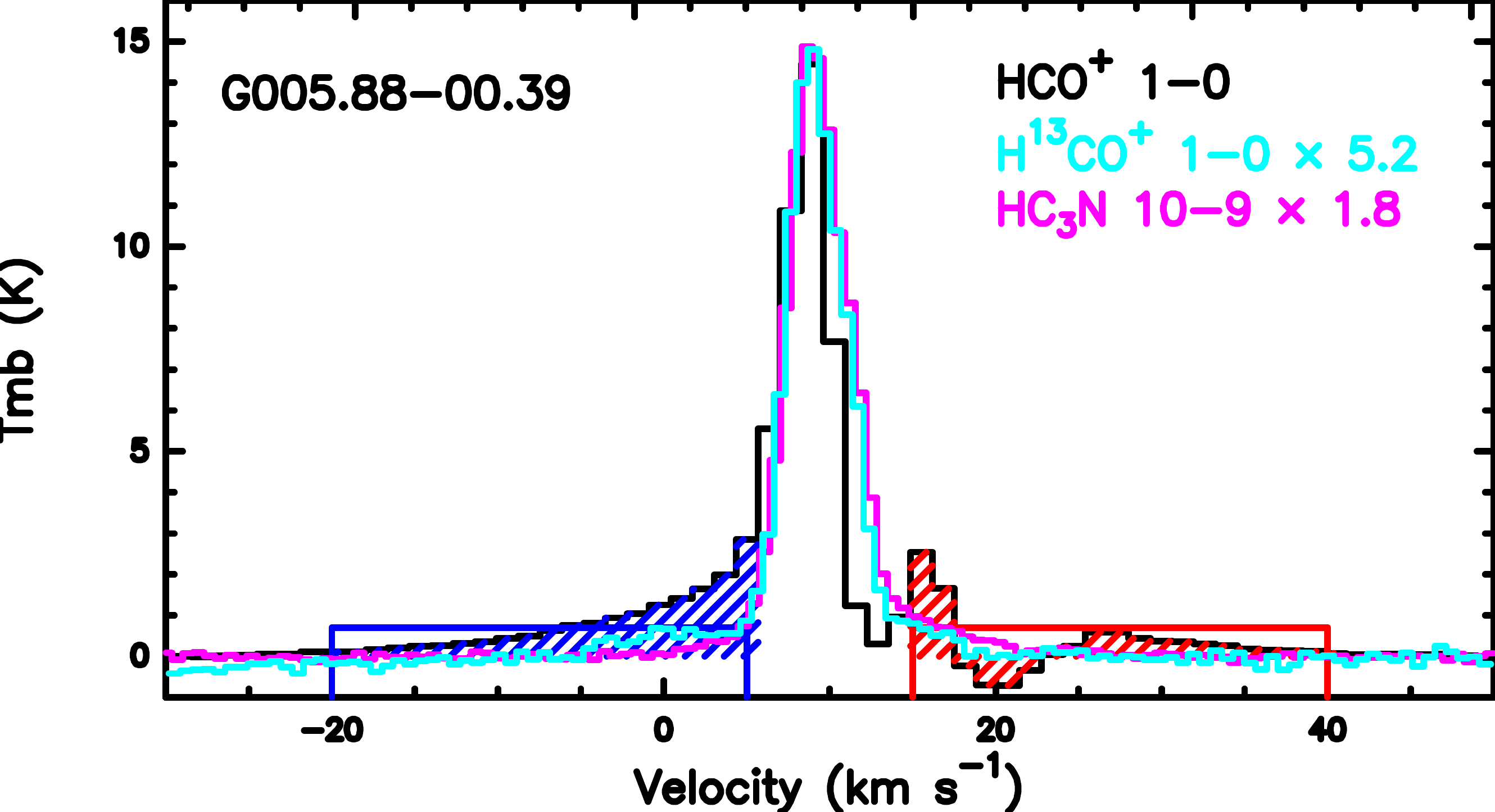}{0.38\textwidth}{(a)}
          \fig{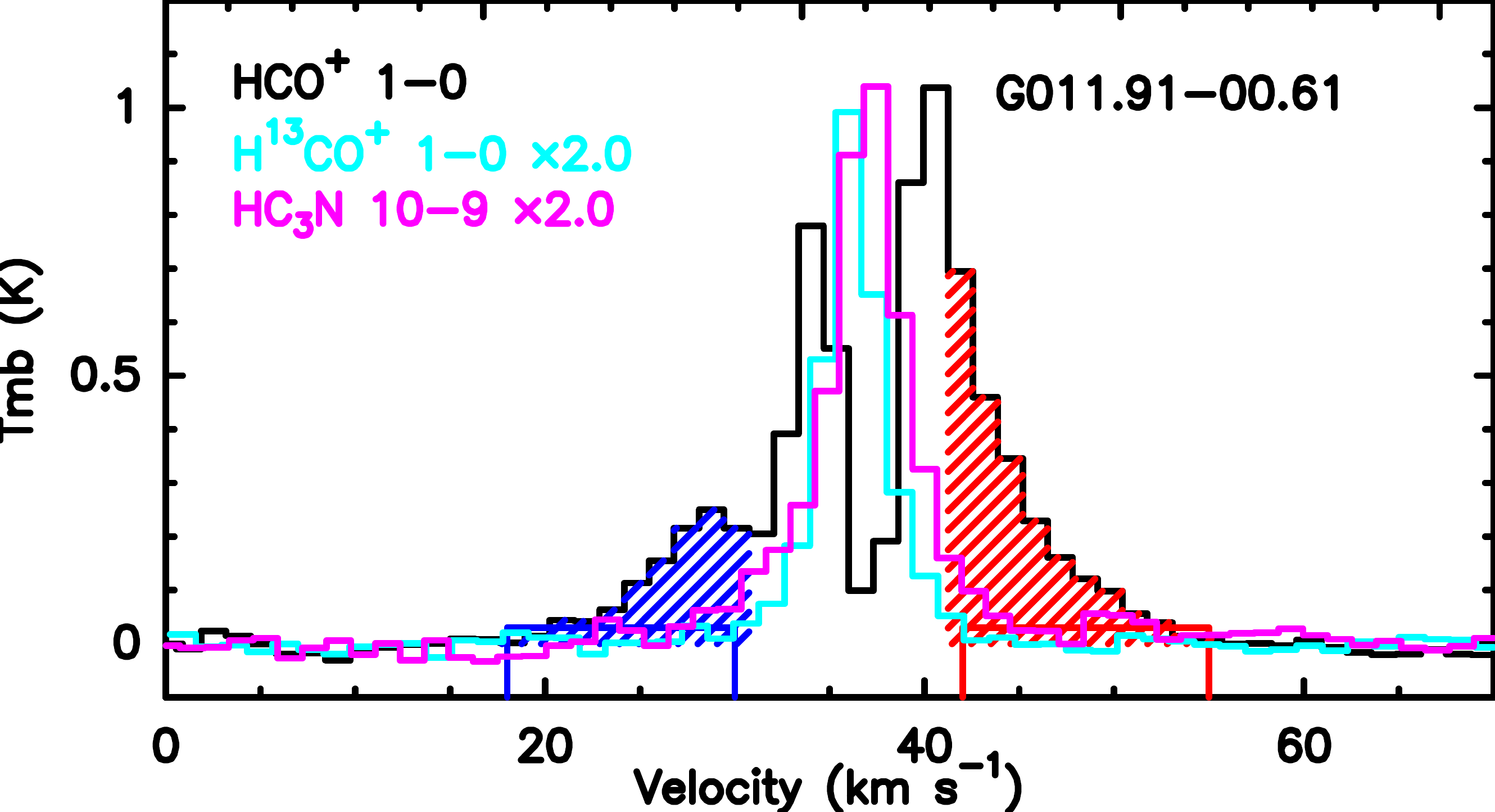}{0.38\textwidth}{(b)}}
\gridline{          
          \fig{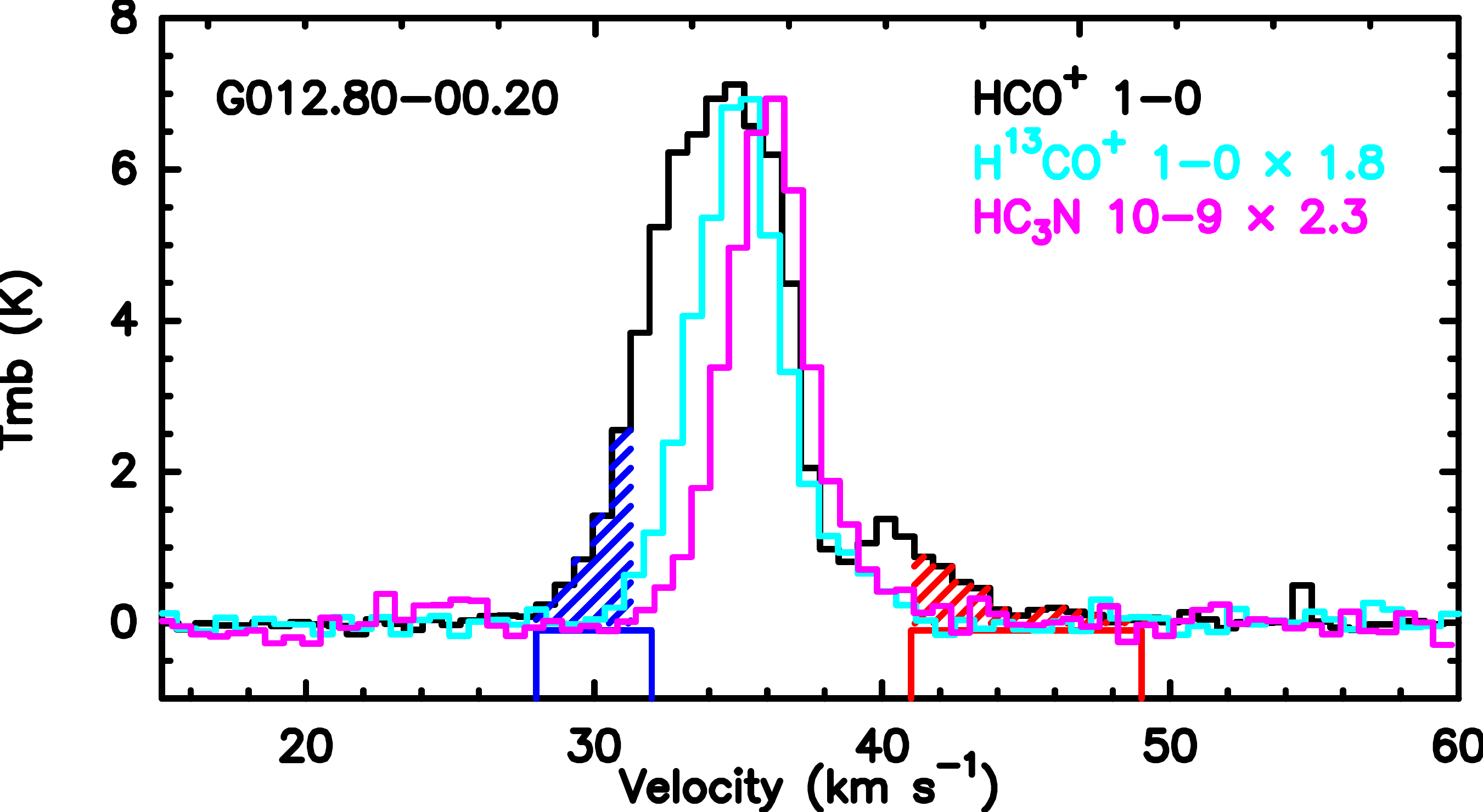}{0.38\textwidth}{(c)}
          \fig{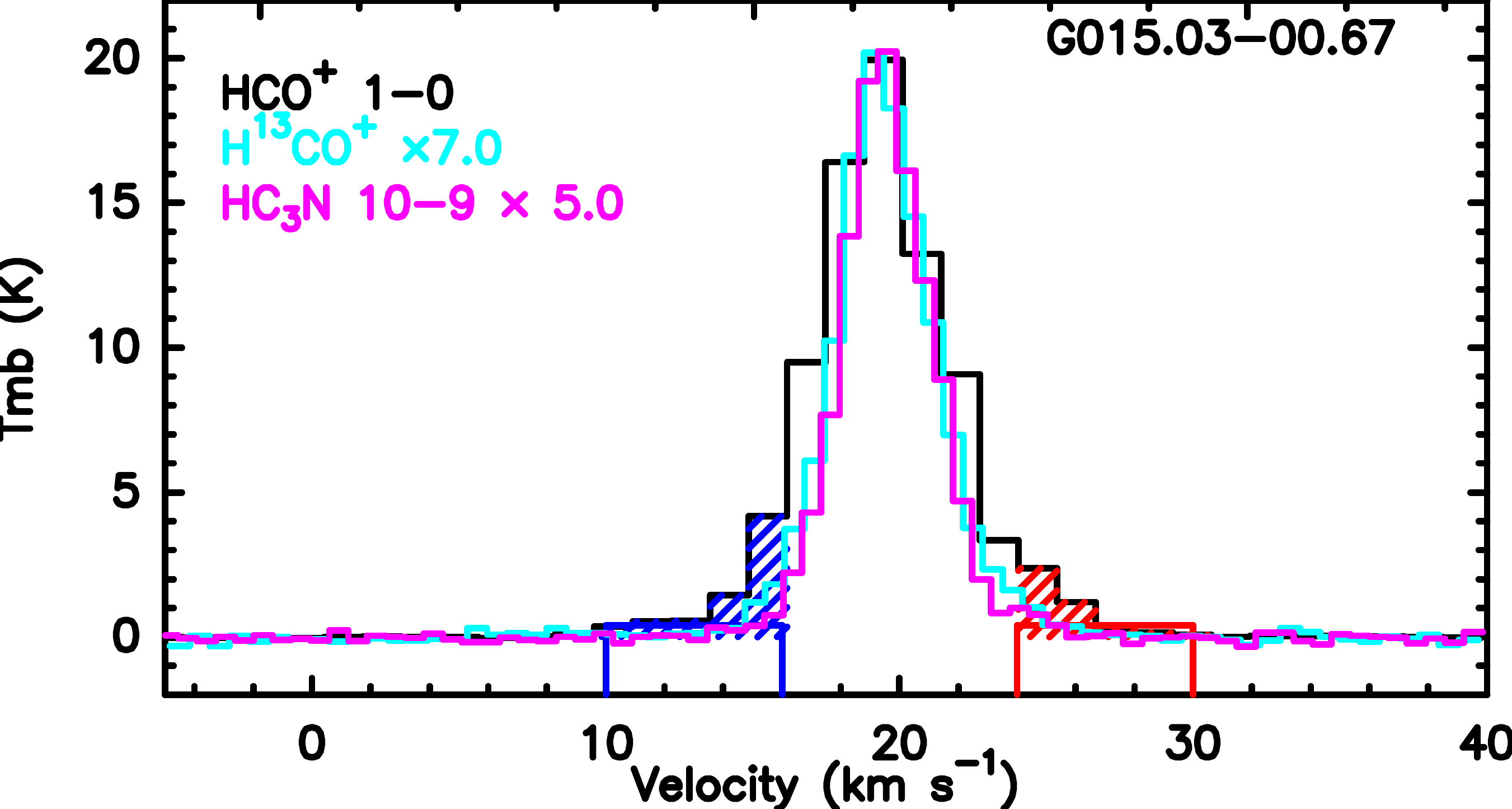}{0.38\textwidth}{(d)}
          }
\gridline{          
          \fig{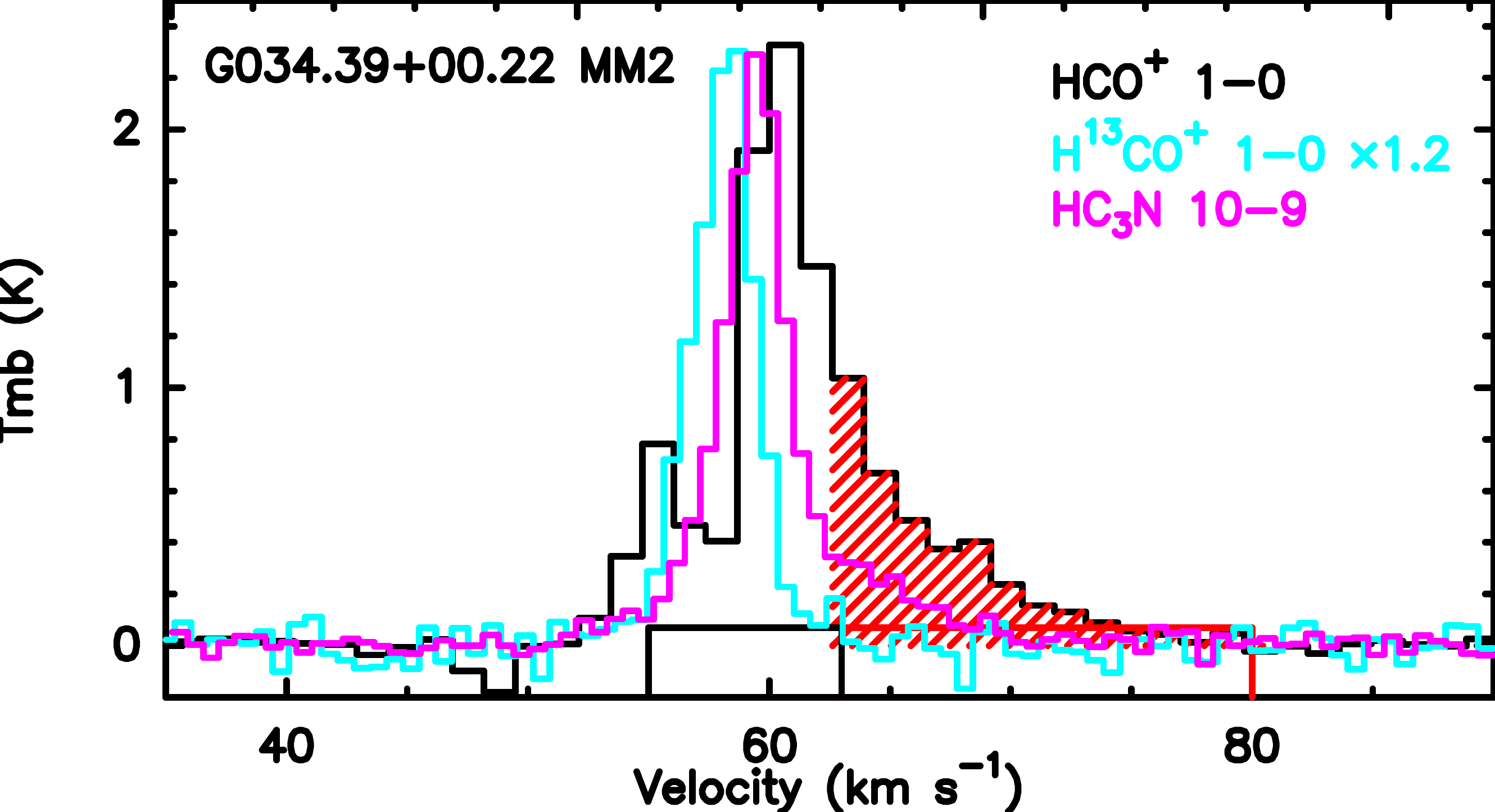}{0.38\textwidth}{(e)}
          \fig{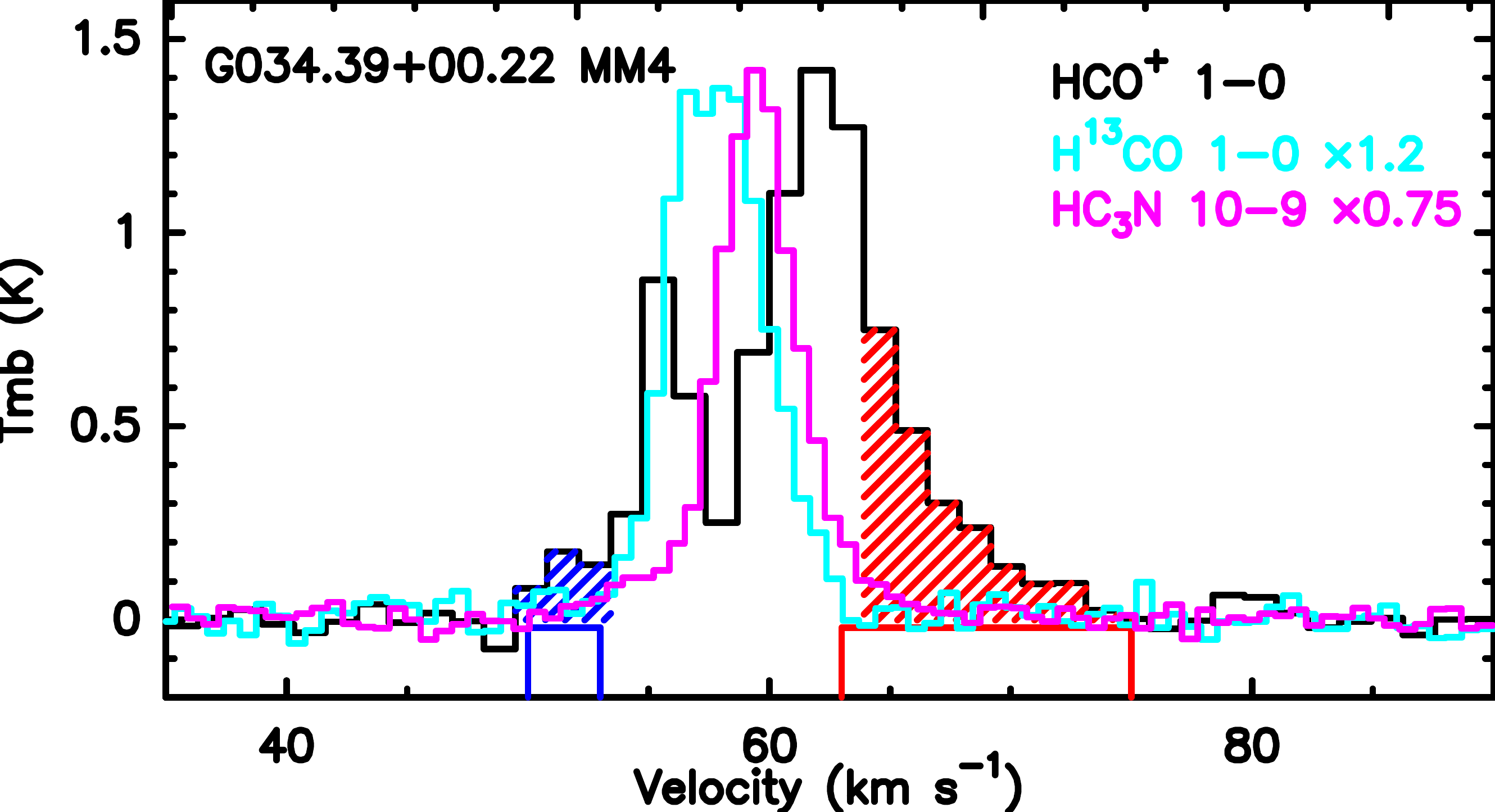}{0.38\textwidth}{(f)}
          }
\gridline{
          \fig{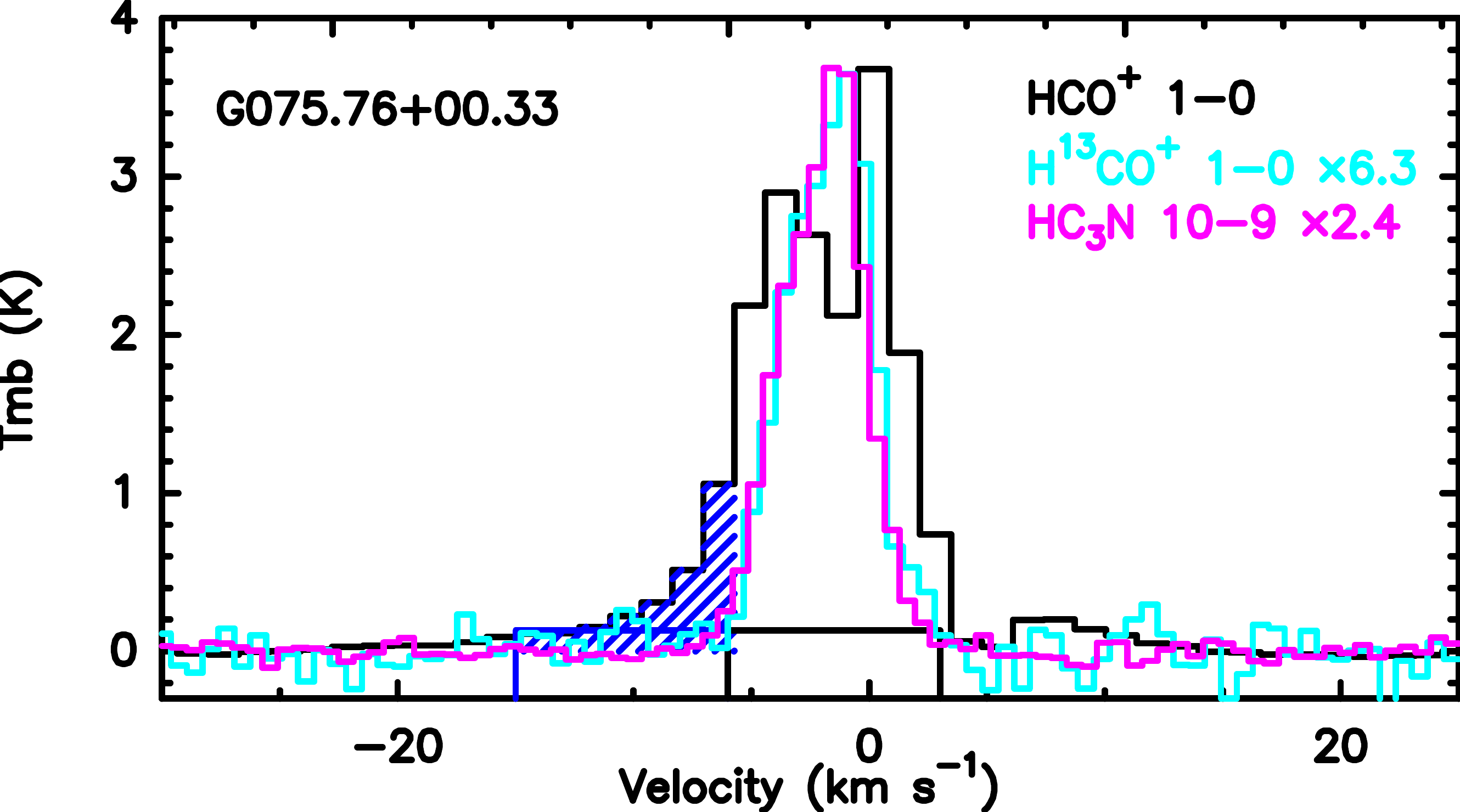}{0.38\textwidth}{(g)}
           \fig{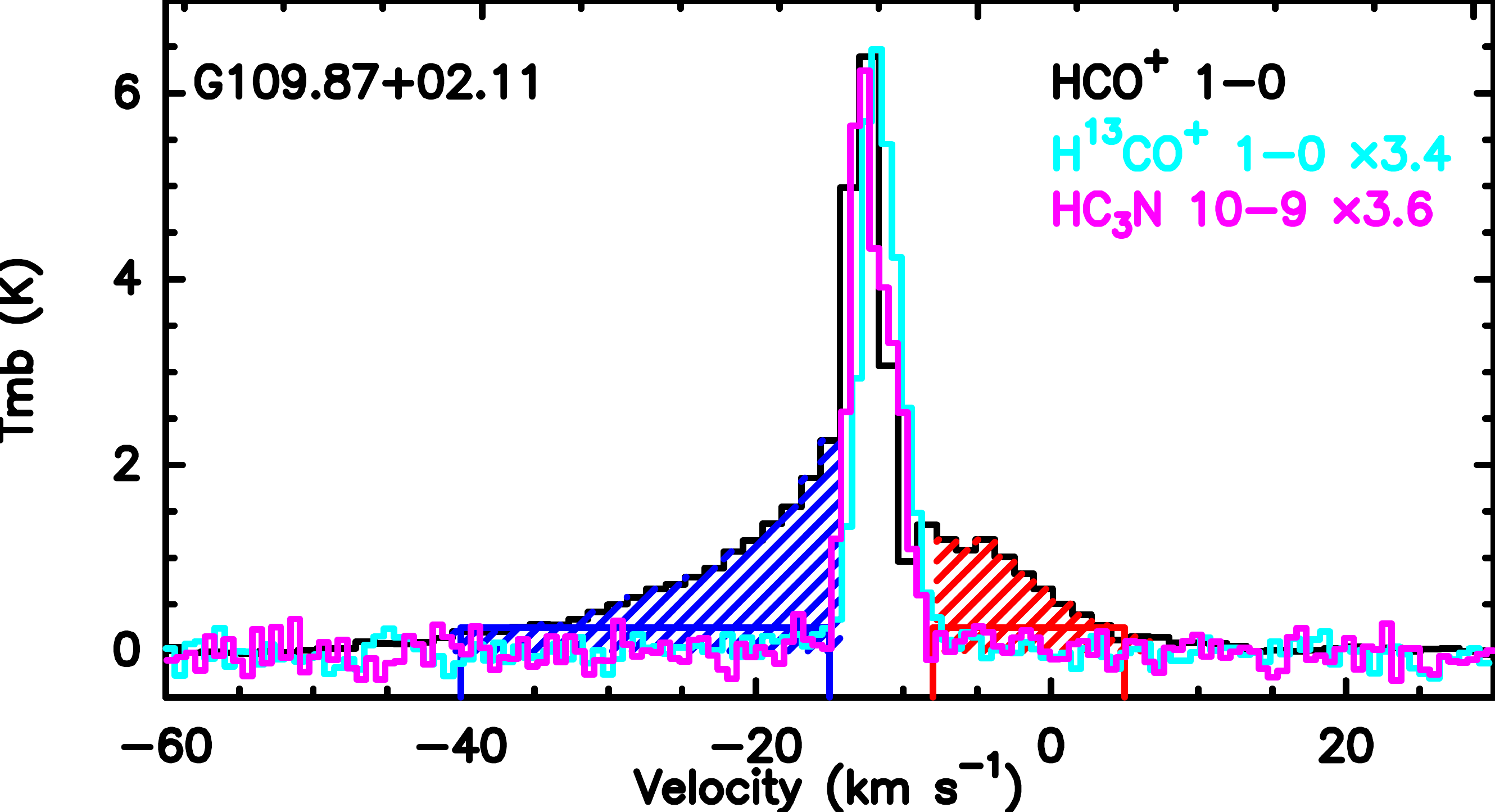}{0.38\textwidth}{(h)}
           }
 \gridline{\fig{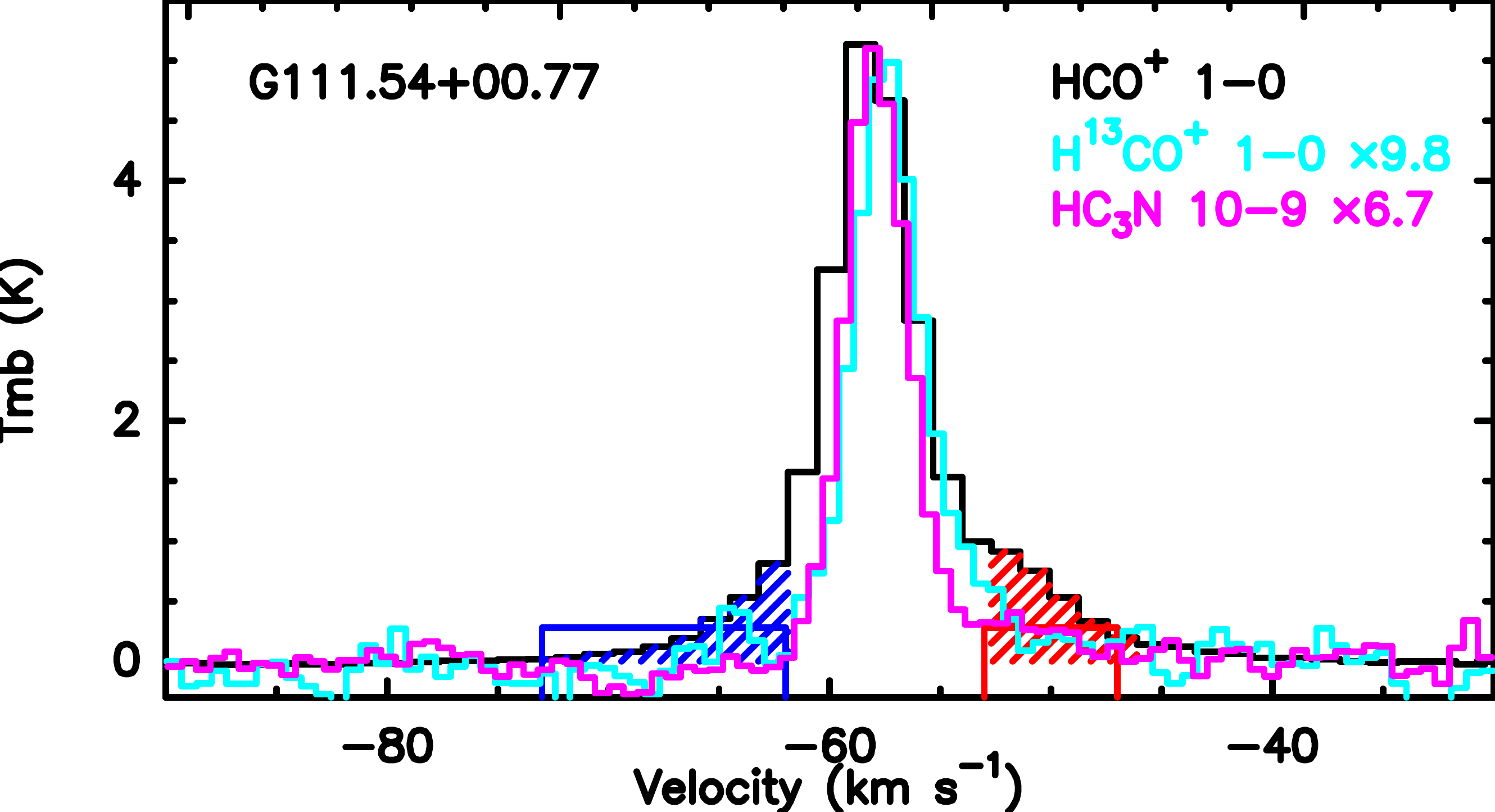}{0.38\textwidth}{(i)}
          \fig{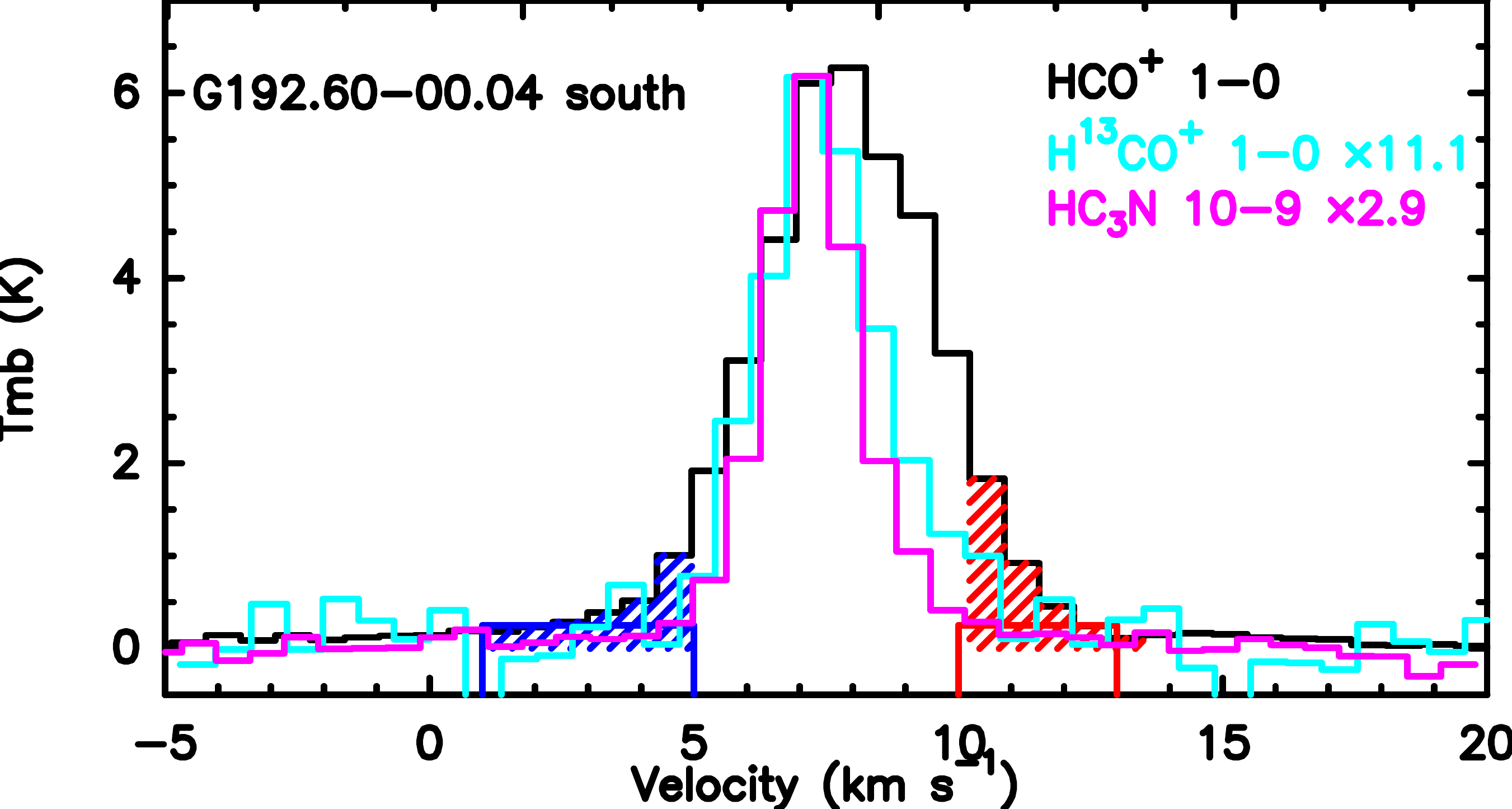}{0.38\textwidth}{(j)}
               }
\caption{The spectra of HC$_{3}$N (10--9) and H$^{13}$CO$^{+}$ (1--0) normalized to the peak intensity of HCO$^+$ (1--0) towards each source. The blue and red line wing emissions are labelled with blue and red windows. The black windows refer to the line core emission ranges. MM2 and MM4 in (e) and (f) refer to the two dust cores identified in \citet{2005rathborne}.}
\label{fig:spec_hcop}
\end{figure*}
\clearpage

\clearpage
\begin{figure*}
\gridline{\fig{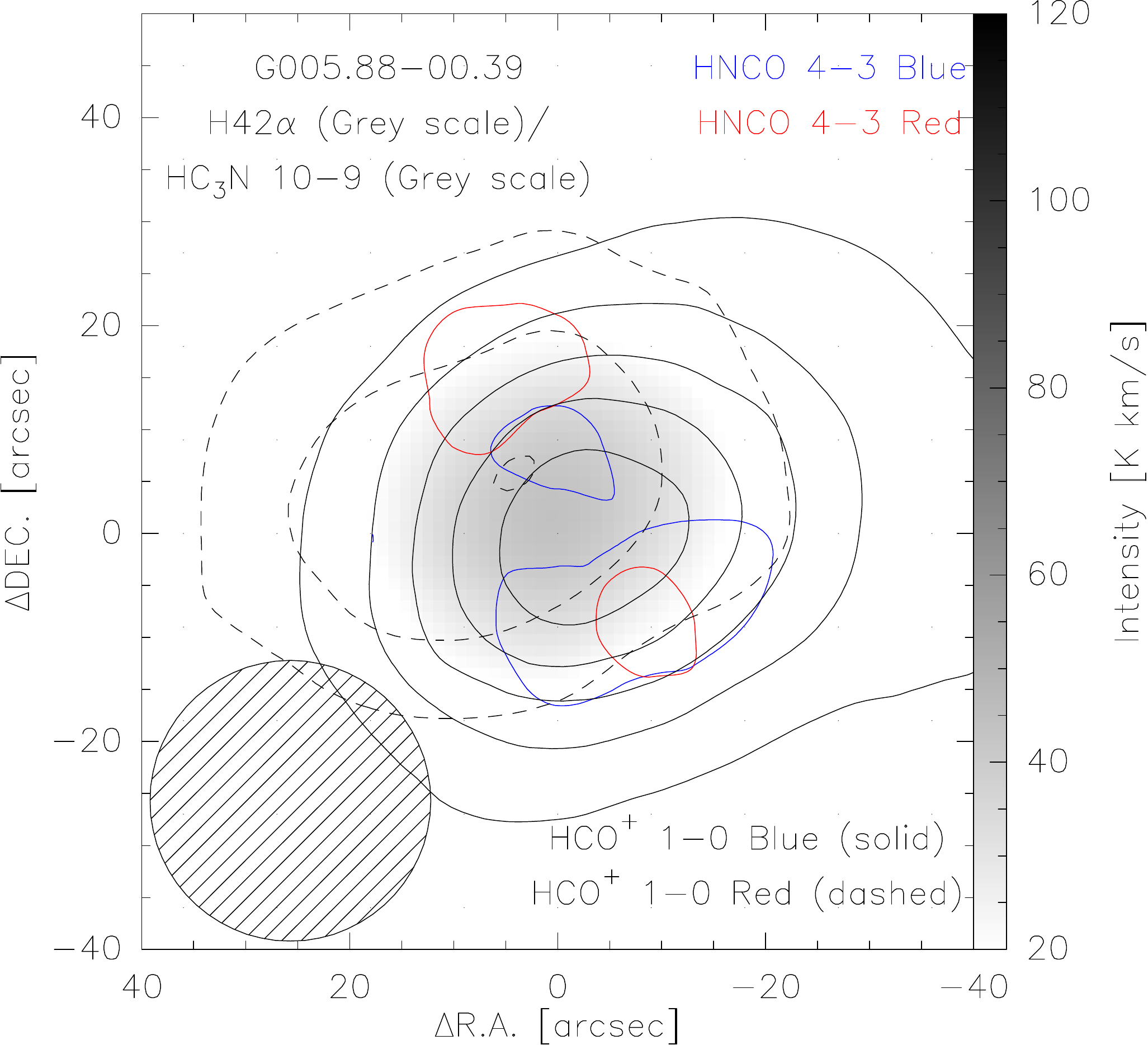}{0.56\textwidth}{(a)}
          \fig{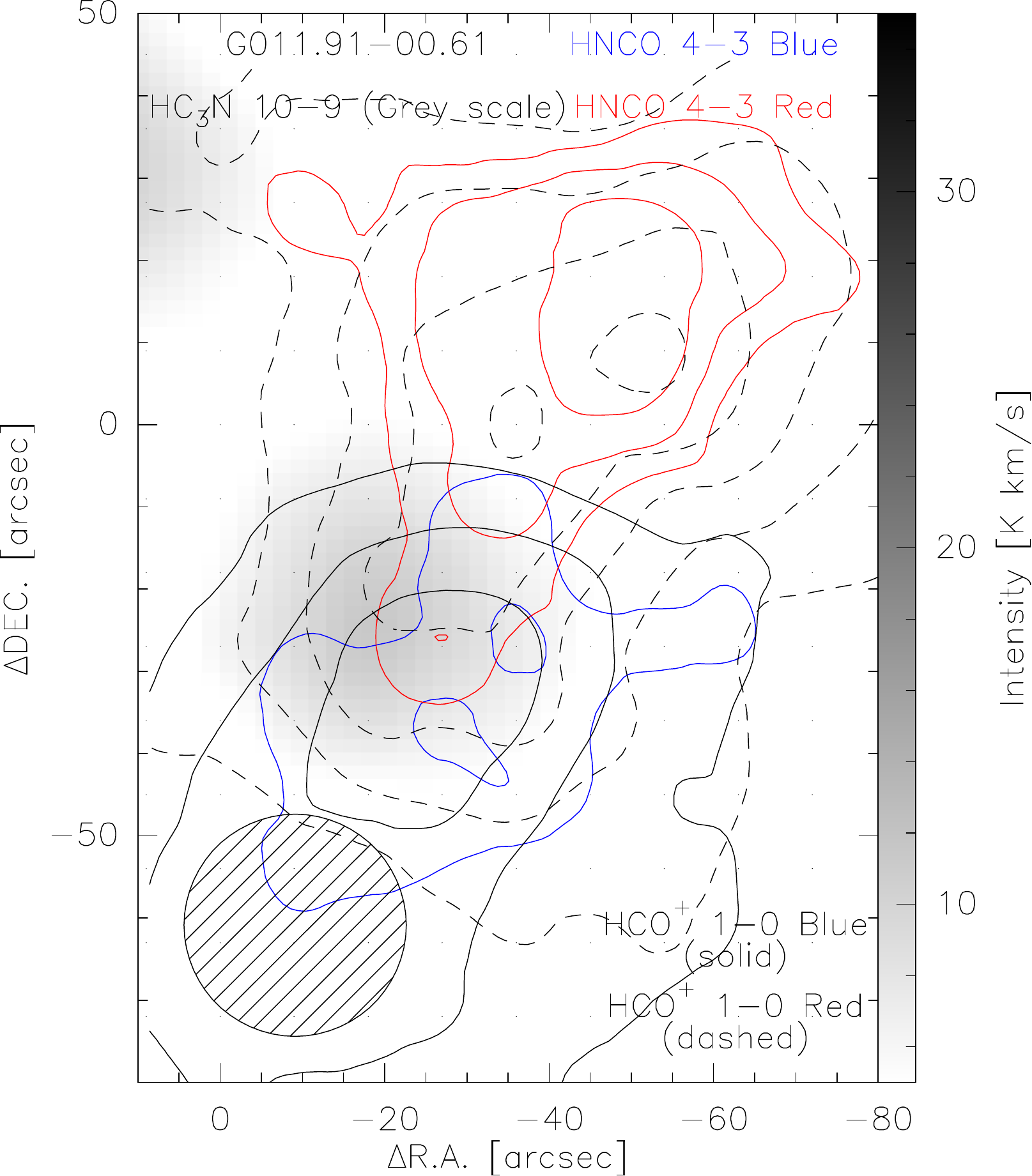}{0.45\textwidth}{(b)}
          }
\gridline{
	\fig{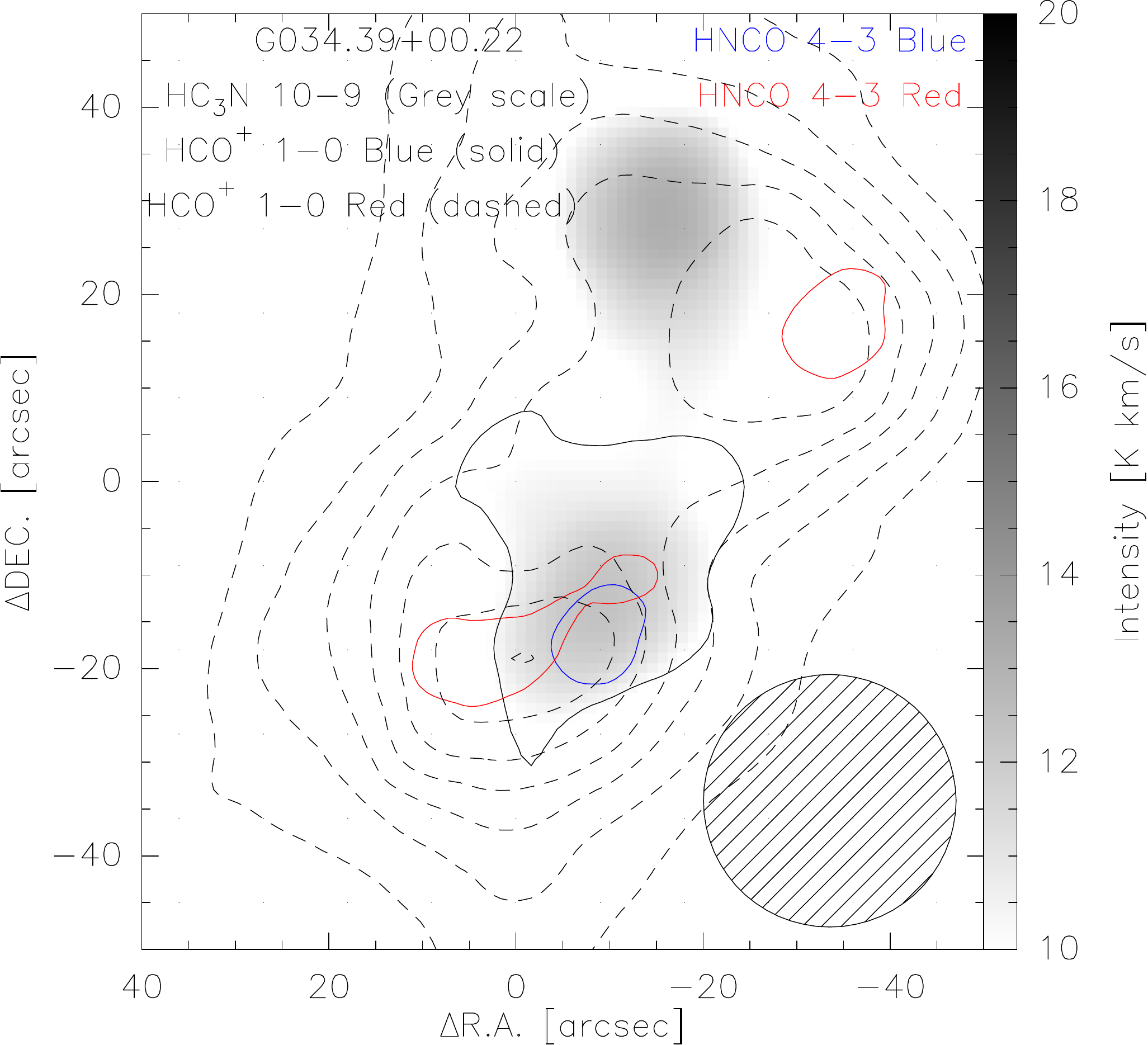}{0.5\textwidth}{(c)}
          \fig{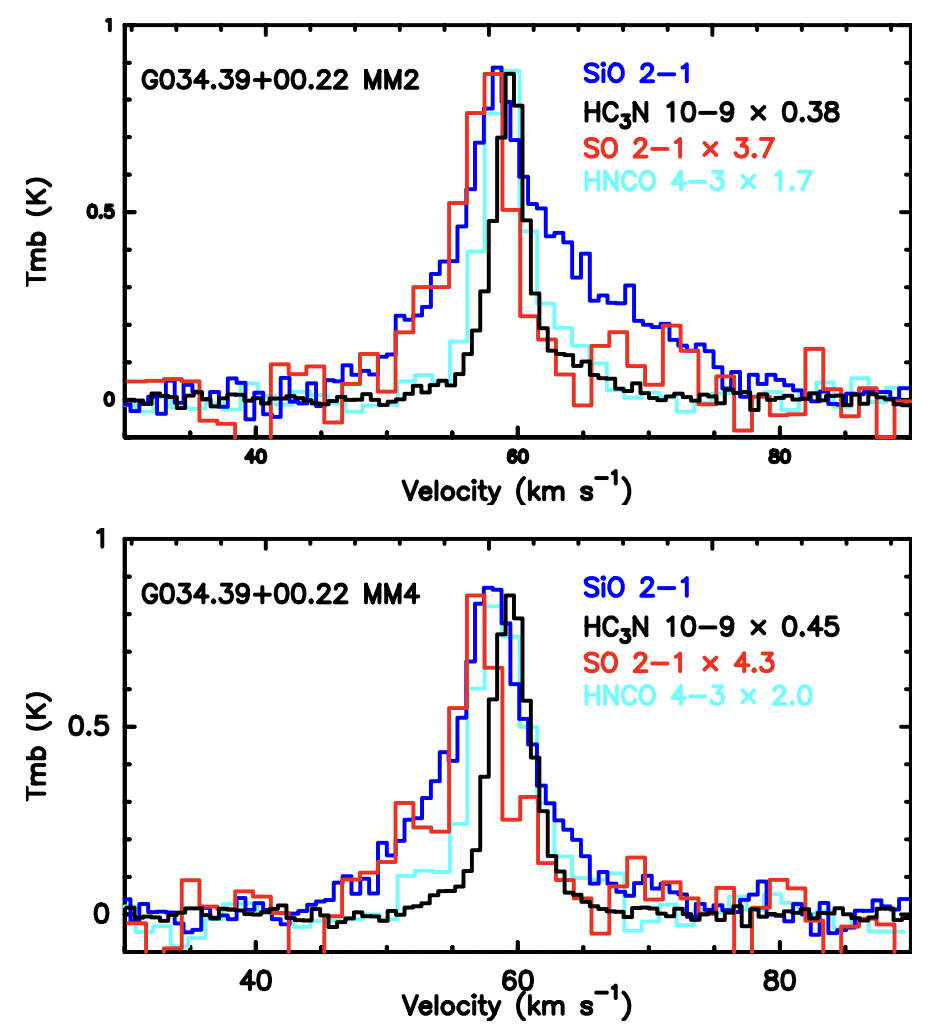}{0.43\textwidth}{(d)}
          }
\caption{The blue and red line wing integrated intensities of HNCO (in blue and red, respectively) and HCO$^+$ (in solid and dashed line, respectively) overlaid with H42$\alpha$ and HC$_{3}$N (in grey scale) for sources labelled as outflow sources. The integrated intensities of HNCO blue and red lobes start from 3$\sigma$ noise level and increase by 2$\sigma$. The integrated intensities of HCO$^+$ blue and red lobes start from 5$\sigma$ noise level and increase by 5$\sigma$, except for G034.39+00.22 where they start from 5$\sigma$ noise level and increase by 2$\sigma$. The beam size is labelled as slash filled circle at the corner of each figure. The greyscale regions represent both H42$\alpha$ and HC$_{3}$N integrated intensities, which have same scale level in source G005.88-00.39 and G011.91-00.61. For G034.39+00.22, the spectra of SO, HNCO and HC$_{3}$N normalized to SiO towards MM2 and MM4 are presented in (d). MM2 and MM4 in (d) refer to the two dust cores identified in \citet{2005rathborne}.  }
\label{fig:outflow}
\end{figure*}

\clearpage

\clearpage
\begin{figure*}
\gridline{\fig{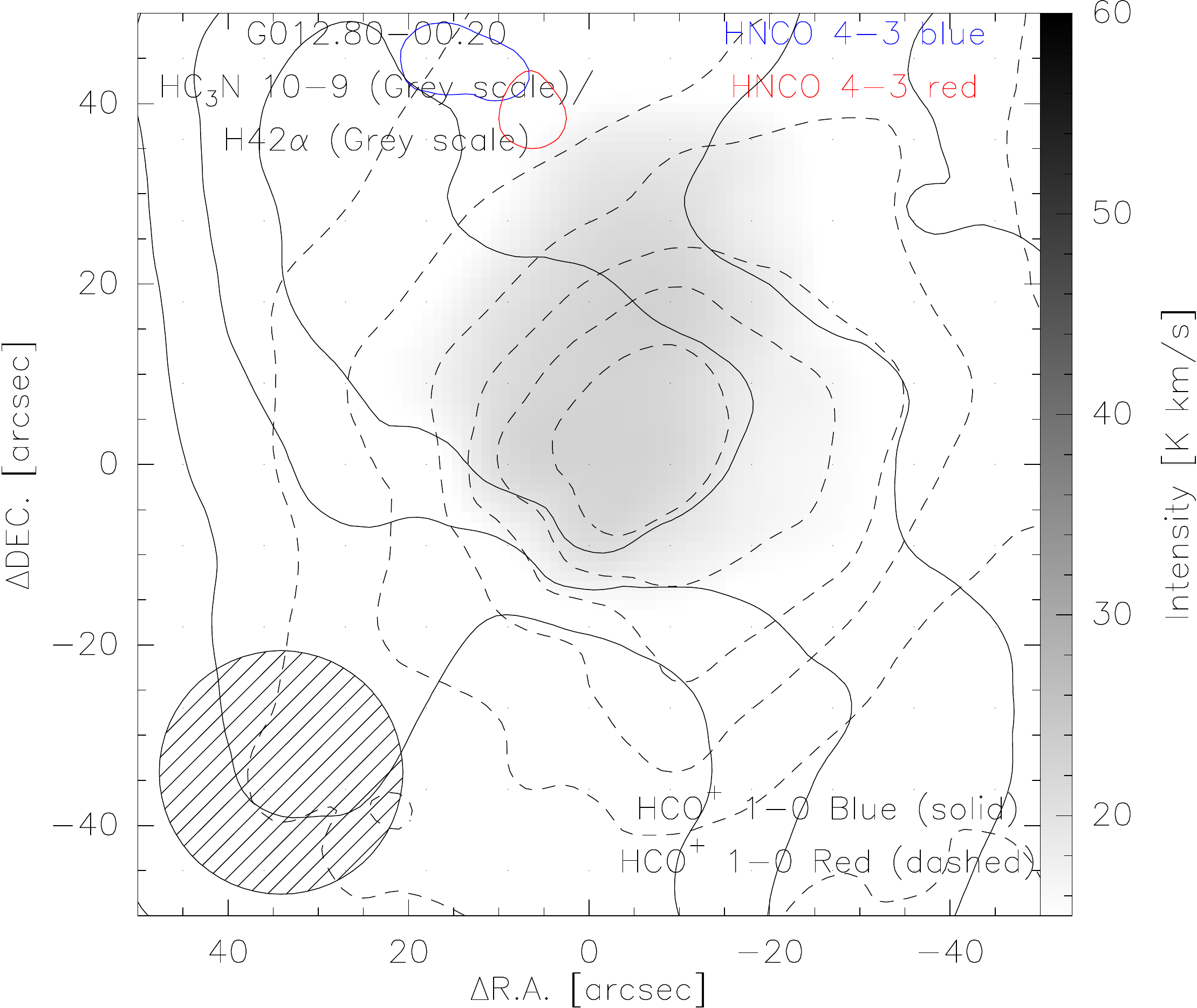}{0.4\textwidth}{(a)}
         \fig{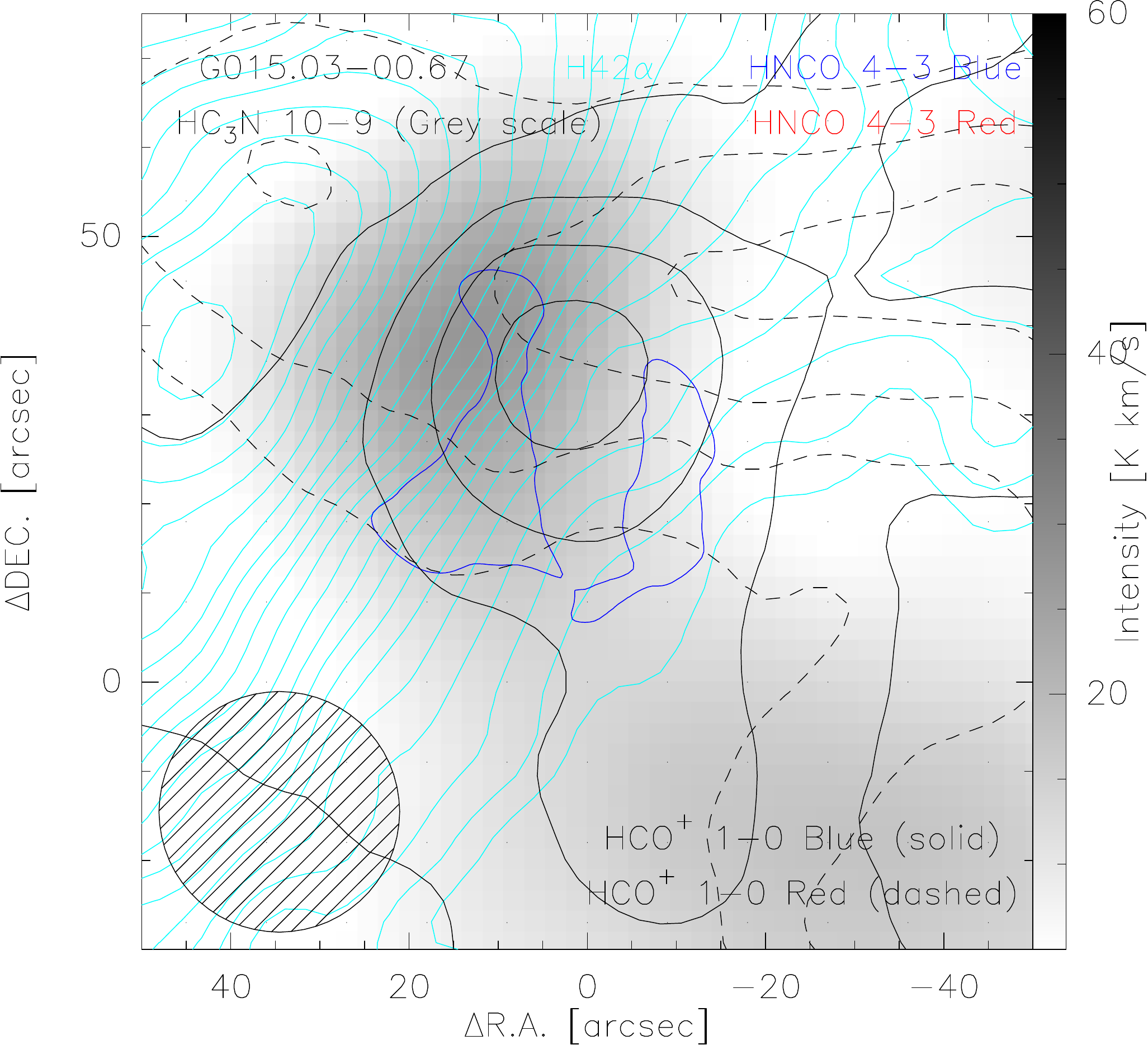}{0.38\textwidth}{(b)}
          }
\gridline{\fig{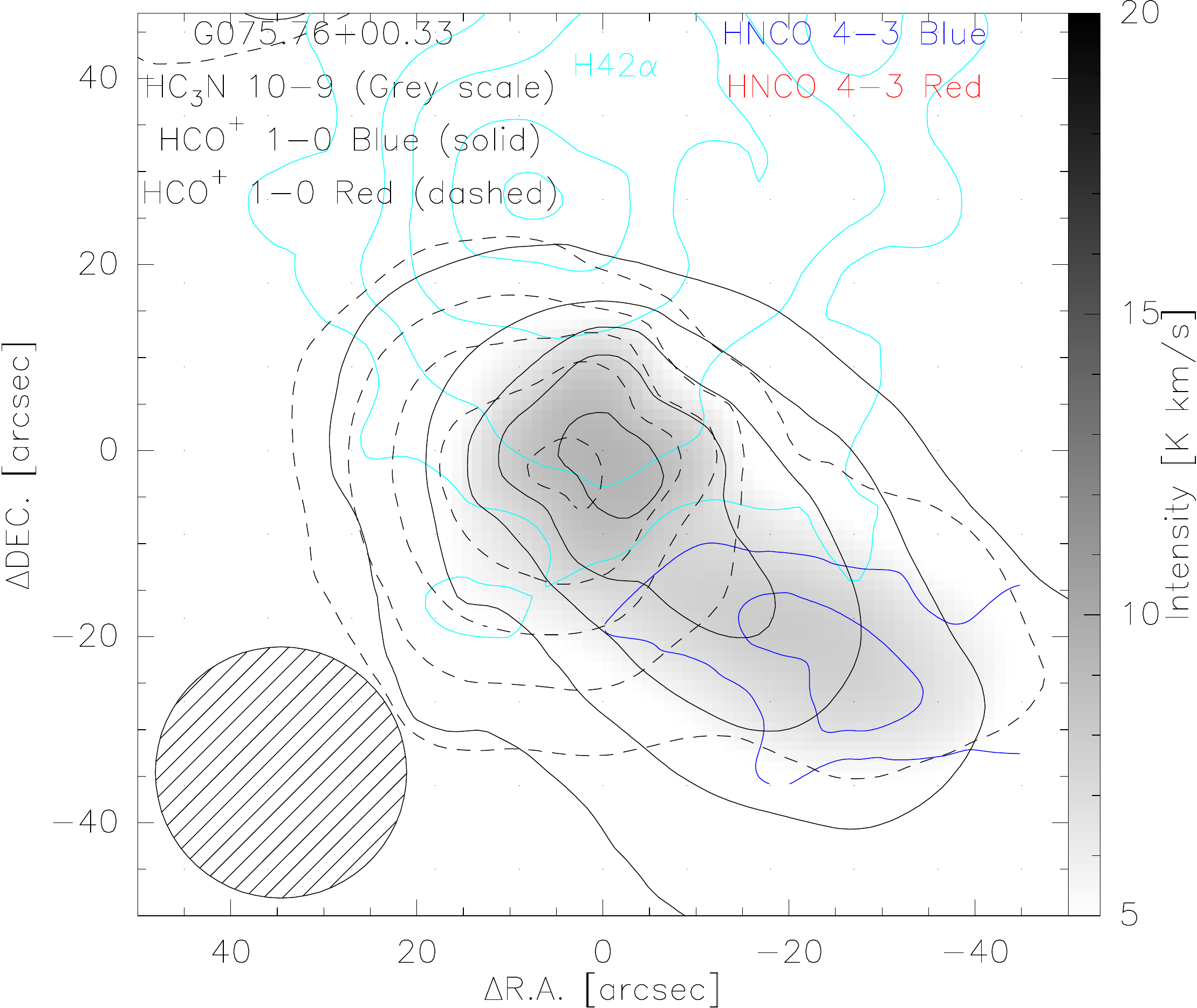}{0.4\textwidth}{(c)}
          \fig{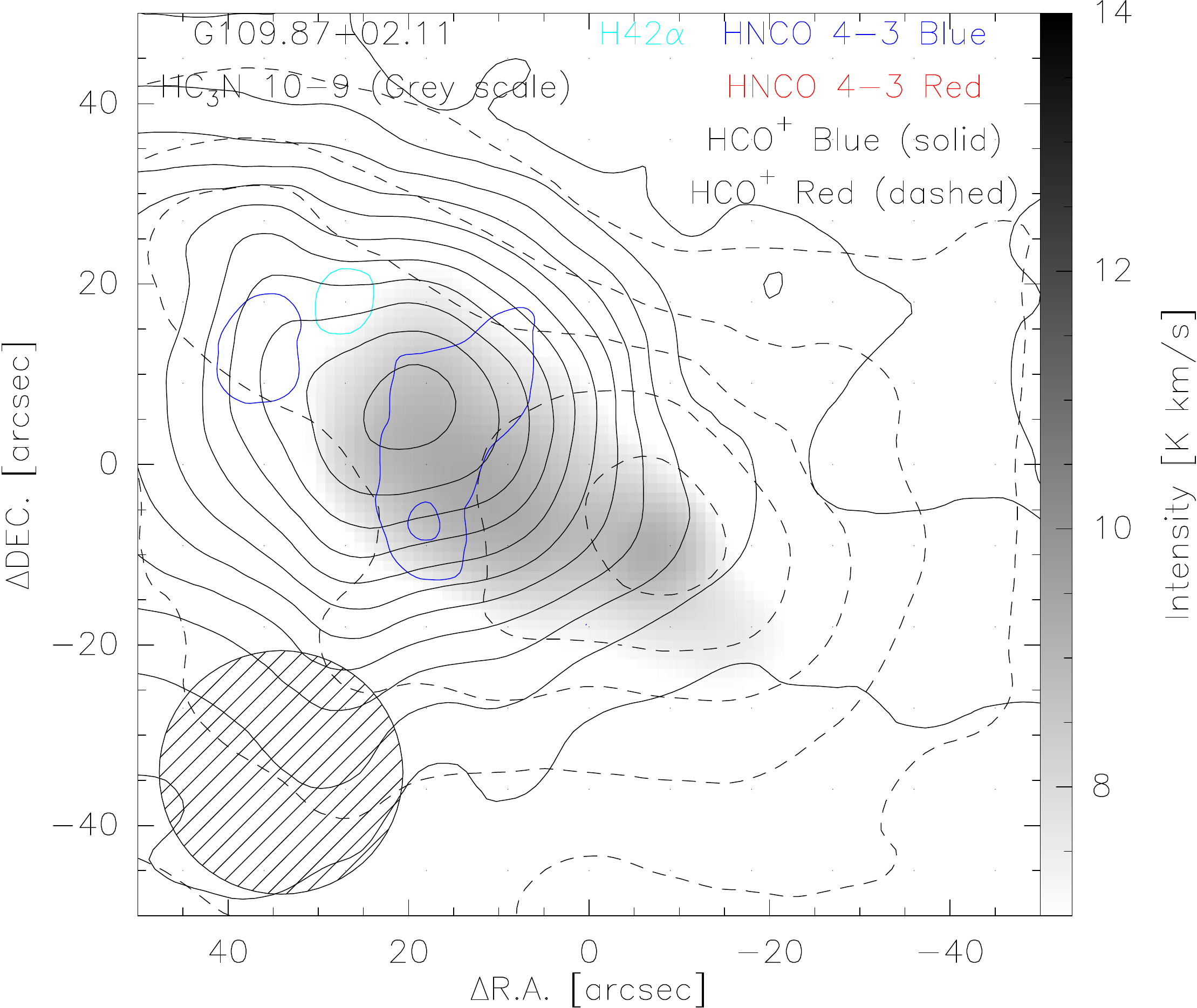}{0.4\textwidth}{(d)}
          }
\gridline{
          \fig{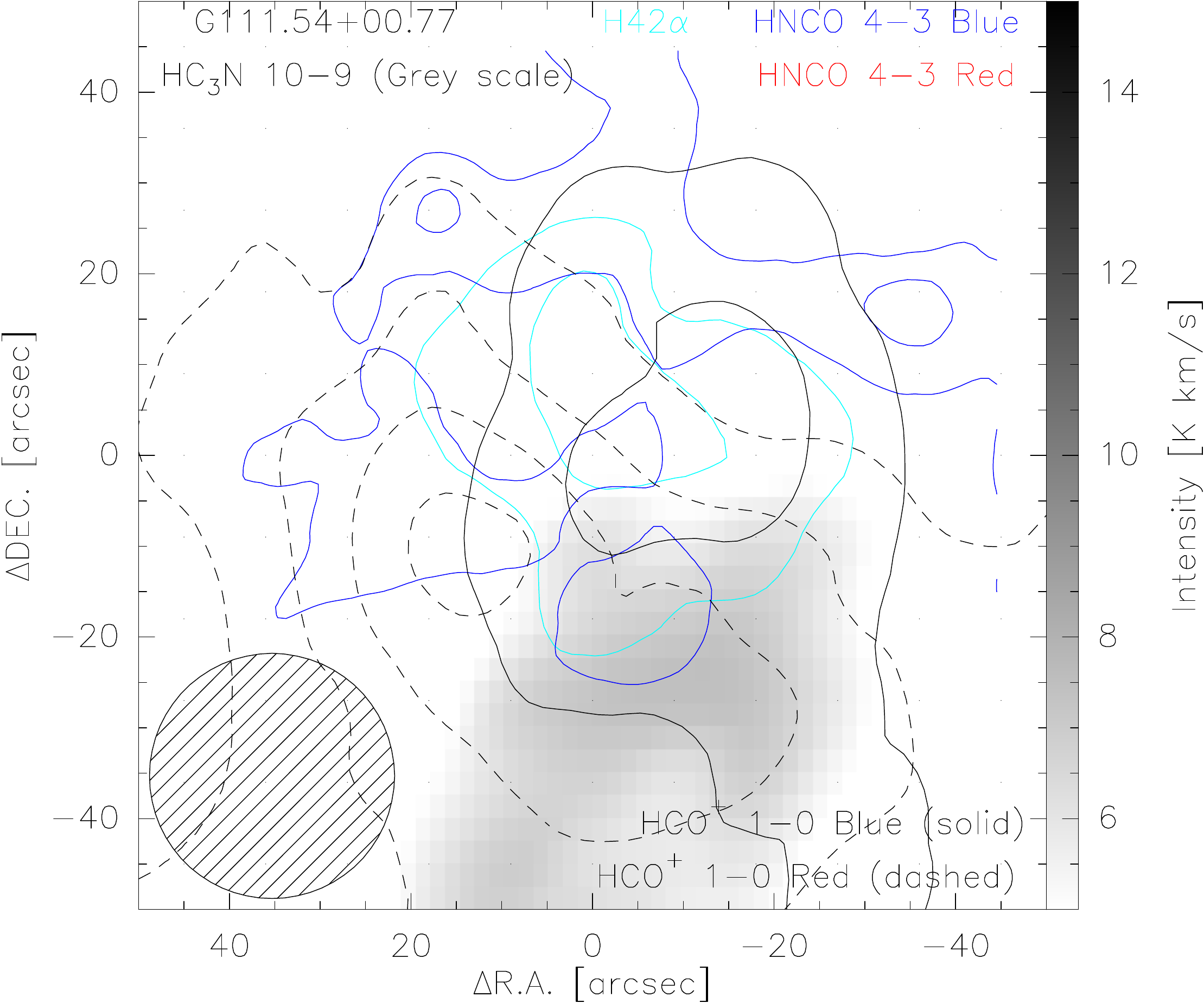}{0.4\textwidth}{(e)}
          \fig{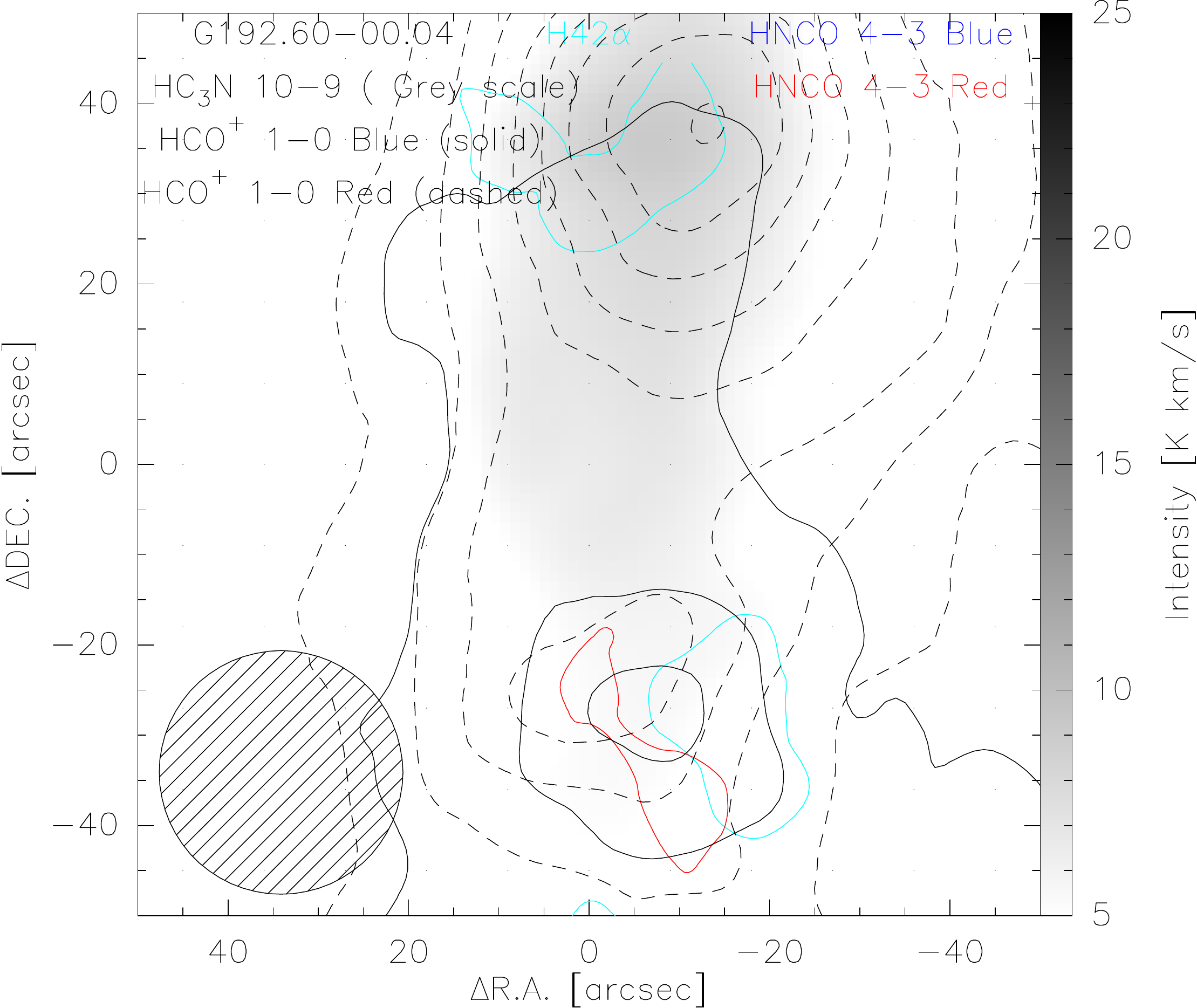}{0.4\textwidth}{(f)}
          }
\caption{The blue and red line wing integrated intensities of HNCO (in blue and red, respectively) and HCO$^+$ (in solid and dashed line, respectively) overlaid with H42$\alpha$ and HC$_{3}$N (in grey scale) for sources labelled as outflow sources. The integrated intensities of HNCO start from 3$\sigma$ noise level and increase by 2$\sigma$. The integrated intensities of HCO$^+$ start from 5$\sigma$ noise level and increase by 5$\sigma$. The contours of H42$\alpha$ integrated intensity start from 5$\sigma$ and increase by 5$\sigma$, except for source G012.80-00.20 where H42$\alpha$ is shown in grey scale as the same level as HC$_{3}$N. The beam size is labelled as slash filled circle at the corner of each figure. }
\label{fig:bluered}
\end{figure*}
\clearpage


\bibliography{HNCOoutflow}{}


\bibliographystyle{aasjournal}



\end{document}